\documentclass[aps,prl,reprint,superscriptaddress,nobibnotes,nofootinbib]{revtex4-2}
\usepackage{graphicx}
\usepackage{dcolumn}
\usepackage{bm}
\usepackage{amssymb}
\usepackage{amsmath}
\usepackage[mathlines]{lineno}
\usepackage{color}
\usepackage{xcolor}

\usepackage{enumerate}
\usepackage{enumitem}
\usepackage{multirow}
\usepackage{chngcntr}
\usepackage{caption}
\usepackage{subcaption}

\let\oldsubequations\subequations
\let\oldendsubequations\endsubequations

\renewenvironment{subequations}
  {\linenomathNonumbers\oldsubequations}
  {\oldendsubequations\endlinenomath}

\begin{document}
\preprint{APS/123-QED}

\title{\textbf{Criticality of nonreciprocal phase oscillators with long-range interactions} 
}

\author{Minwoo Bae}
\email{Contact author: minwoo-bae@g.ecc.u-tokyo.ac.jp}
\affiliation{
 Department of Complexity Science and Engineering, Graduate School of Frontier Sciences, the University of Tokyo, Chiba 277-8561, Japan
}

\author{Hiroshi Kori}
\affiliation{
 Department of Complexity Science and Engineering, Graduate School of Frontier Sciences, the University of Tokyo, Chiba 277-8561, Japan
}

\date{\today}

\begin{abstract}
We study noisy identical Kuramoto--Sakaguchi oscillators with phase lag $\alpha\in[0,\pi/2)$, where $\alpha>0$ induces nonreciprocal interactions. For long-range interactions with a power-law decay controlled by a parameter $\sigma$ on $d$-dimensional lattices for $d=1$ and 2, numerical phase diagrams in the \((\sigma,\alpha)\) plane reveal a critical phase lag $\alpha_c$, below which spontaneous synchronization emerges and above which nonreciprocity becomes relevant. The critical behavior near $\alpha=\alpha_c$ belongs to a universality class distinct from that of the long-range Edwards-Wilkinson fixed point. We characterize the transition analytically using dynamical renormalization group theory. 
\end{abstract}

\maketitle
\textit{Introduction--} Nonreciprocal interactions have emerged as a major topic in the study of nonequilibrium phenomena in complex systems~\cite{fruchart2026nonreciprocal-review}. Unlike reciprocal interactions, they break the symmetry between the interactions of two components. Such nonreciprocity affects the properties of active matter~\cite{active-Fruchart2021Nonreciprocal,active-sync-PRL}, the steady states of biological and social systems~\cite{Strogatz-conformist-contrarian}, dynamical pattern formation in diffusive systems~\cite{active-generic-route}, the emergence of dynamical frustration~\cite{Nonreciprocal-PRX}, and critical behavior in spin systems~\cite{Nonreciprocal-Isin-PRL,quenched-noreciprocal-ferromagnet,Nonreciprocal-XY}. The \textit{relevance} of nonreciprocity to phase transitions has also been investigated in short-range interacting systems with at least two coupled fields~\cite{when-is-nonreciprocity-relevant}. Thus, the role of nonreciprocity in critical phenomena has been explored in both spin models and dynamical systems.

In this regard, we are led to investigate how the interplay between nonreciprocity and long-range interactions shapes critical behavior, as the criticality of systems with independently tunable nonreciprocity and interaction range remains largely unexplored. We address this question using the Kuramoto--Sakaguchi model~\cite{Kuramto-Sakaguchi}, where phase lag $\alpha$ explicitly induces nonreciprocal interactions~\cite{pnas-nonreciprocal,Nonreciprocal-PRX}. Phase lag is not merely a formal extension of the Kuramoto oscillator~\cite{kuramoto1975self,kuramoto1984chemical}; it has been used to describe spiral patterns~\cite{Pattern-Spiral-Kuramoto,Spiral-Chimera}, chimera states~\cite{Spiral-Chimera,Chimera-Storgatz-2004,Chimera-Topological-defect}, and twisted states on lattices~\cite{QTwisted_Lee}, and has also been implemented in experimental oscillator systems~\cite{Experiment-circuit,Kiss-Rusin-Kori-John}. Thus, the Kuramoto--Sakaguchi model provides a minimal framework for elucidating how nonreciprocity influences critical behavior beyond the short-range interactions.


In this Letter, we characterize the criticality of the noisy Kuramoto--Sakaguchi model with identical oscillators and long-range interactions on $d$-dimensional lattices. Denoting the distance between nodes $i$ and $j$ by $r_{ij}$, we consider an interaction strength that decays as $(r_{ij}^2+r_0^2)^{-(d+\sigma)/2}$, where $r_0$ is a finite constant. Our main results are as follows. We identify a critical phase lag \(\alpha_c\), below which spontaneous synchronization emerges. The threshold \(\alpha_c\) decreases monotonically with increasing \(\sigma\); synchronization is more difficult for shorter interaction ranges. In this regime, nonreciprocity is irrelevant in the renormalization-group sense: under coarse graining, the renormalized phase lag flows to zero, and the synchronized phase is governed by an infrared (IR) fixed point corresponding to the long-range Edwards--Wilkinson (EW) universality class~\cite{EW-original,long-range-EW}. By contrast, for \(\alpha>\alpha_c\), nonreciprocity becomes relevant and destroys spontaneous synchronization.
The critical behavior near \(\alpha=\alpha_c\) is nontrivial and distinct from that associated with the long-range EW fixed point. Using dynamical renormalization group (DRG) theory, we derive the critical properties governing the transition and the associated critical exponents. These predictions are quantitatively confirmed by finite-size scaling analysis of numerical simulations.

\textit{Model--} We investigate noisy identical Kuramoto--Sakaguchi model on $d$-dimensional Eulidean lattices: 
\begin{equation} \frac{d\theta_i}{dt} = K \sum_{j=1}^N J_{ij} \sin(\theta_j-\theta_i-\alpha) + \eta_i(t), \label{eq:model} \end{equation} where $\theta_i$ is the phase of the oscillator at node $i$, $K>0$ is the coupling strength, $J_{ij}=(r_{ij}^2+r_0^2)^{-(d+\sigma)/2}/\sum_j (r_{ij}^2+r_0^2)^{-(d+\sigma)/2}$ is the normalized long-range weight between nodes $i$ and $j$, and $\alpha\in[0,\pi/2)$ is the phase lag. $\eta_i(t)$ is the white Gaussian noise with zero mean, satisfying \begin{equation} \langle \eta_i(t)\eta_j(t’) \rangle = D\delta_{ij}\delta(t-t’), \label{eq:eta-D} \end{equation} where $D>0$ is the noise strength. 
We set $K=1$ without loss of generality by rescaling time $t$ and noise strength $D$. We focus on $\alpha\ge0$, because the dynamics for $\alpha$ and $-\alpha$ are statistically equivalent under $\theta_i\to-\theta_i$, together with the sign reversal of the noise realization.

For $\alpha=0$, Eq.~(\ref{eq:model}) satisfies detailed balance, and its stationary distribution coincides with the Gibbs distribution of the XY model at temperature $T=D/2$~\cite{RevModPhys-Kuramoto}.
For $\alpha\ne0$, by contrast, the dynamics is then no longer of gradient form, and the steady state cannot generally be described by an equilibrium Gibbs measure. 

\begin{figure}[t]
    \centering
    \includegraphics[width=1\linewidth]{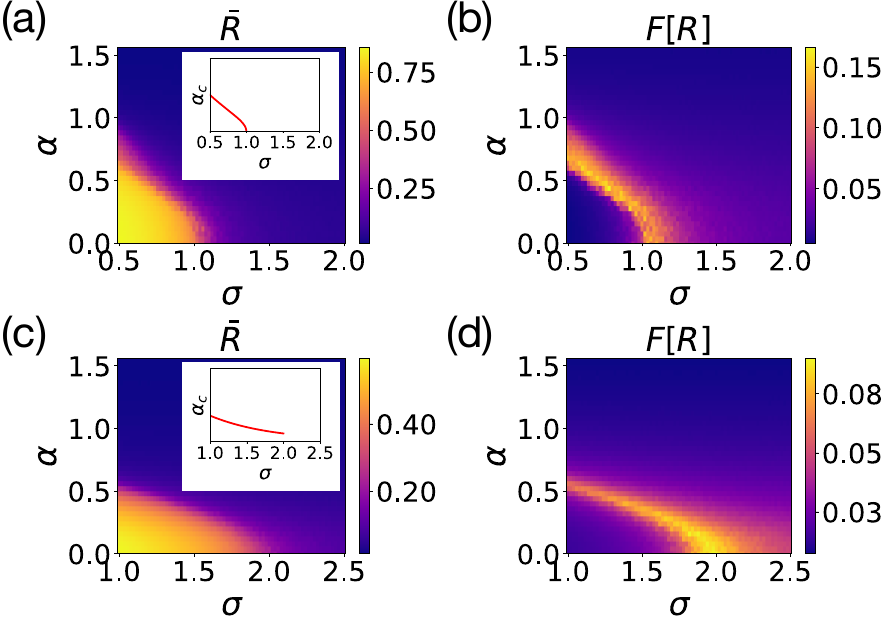}
    \captionsetup{justification=raggedright,singlelinecheck=false}
    \caption{Phase diagram of Eq.~(\ref{eq:model}) in the $(\sigma,\alpha)$ plane. The initial phases $\theta_i$ are independently drawn from a uniform distribution. Panels (a) and (b) correspond to $d=1$ with $D=0.36$, whereas panels (c) and (d) correspond to $d=2$ with $D=0.7225$ (see End Matter). The system size is $N=2^{12}$ for all panels, and $r_0=0.2$. The color bars show the ensemble-averaged values over 5 noise and initial-phase realizations of (a,c) the long-time-averaged synchronization order parameter $\bar{R}$ and (b,d) the dynamical fluctuation $F[R]$ [Eq.~(\ref{eq:F[r]})], obtained by numerical integration with  $t_r=1.8\times10^5$ and $t_{\mathrm{avg}}=2\times10^4$. Inset: analytical dependence of the critical phase lag $\alpha_c$ on $\sigma$ obtained from Eqs.~(\ref{eq:alpha_c-text}) and (\ref{eq:G-coeff-text}), with $r_0=0.2$ and $\Lambda=1$.}
    \label{fig:1}
\end{figure}

\textit{Numerical phase diagram--} In our numerical analysis, we measure
the synchronization order parameter
\begin{equation}
    R=\frac{1}{N}\left|\sum_{j=1}^N e^{i\theta_j}\right|,
    \label{eq:r}
\end{equation}
and its dynamical fluctuation,
\begin{equation}
    F[R]= \sqrt{\frac{1}{t_{\mathrm{avg}}}\int_{t_r}^{t_r+t_{\mathrm{avg}}} \left[R(t)-\bar{R}\right]^2 dt},
    \label{eq:F[r]}
\end{equation}
where $\bar{R}=\frac{1}{t_{\mathrm{avg}}}\int_{t_r}^{t_r+t_{\mathrm{avg}}} R(t)\ dt$
is the time average of $R$ measured after discarding an initial transient ($t_r>0$). We numerically integrate Eq.~(\ref{eq:model}) from random initial phases $\theta_i$ using the fourth order Runge--Kutta method with time step $\Delta t=0.1$. 

Figure~\ref{fig:1} presents numerical results in the weak-noise regime (small $D$). Because finite-size phase fluctuations are controlled by the linear system size $L$ rather than the number of nodes $N$, we use different noise strengths $D$ for $d=1$ and $d=2$; the criterion for this choice is provided in End Matter. As shown in Figs.~\ref{fig:1}(a) and \ref{fig:1}(c), for $\alpha=0$, the synchronization transition occurs at $\sigma=\sigma_c$. For $d=1$ and $d=2$, the thermodynamic-limit threshold $\sigma_c=d$ follows from the Mermin--Wagner theorem, as generalized by Cassi to long-range interactions via the spectral dimension $d_s$~\cite{Mermin-Wagner,Mermin-classical,Cassi-MerminWagner} (see End Matter), with \textit{spontaneous} synchronization emerging for $\sigma<\sigma_c$~\cite{Inverse-Mermin-Wagner-Burioni}. However, for $\alpha>0$, we find that synchronization persists only up to a finite critical phase lag $\alpha_c$. Remarkably, $\alpha_c$ exhibits a clear dependence on $\sigma$, and synchronization is lost for $\alpha>\alpha_c$. In Figs.~\ref{fig:1}(b) and \ref{fig:1}(d), the enhancement of $F[R]$ (bright colors) near $\alpha_c$ indicates the instability of the incoherent state, suggesting a continuous phase transition. 

\textit{Dynamical Renormalization Group--} In the following, we analytically investigate the critical behavior. We first note that the linearized form of Eq.~(\ref{eq:model}), obtained under the assumption of small phase-difference $|\theta_i-\theta_j|\ll1$, cannot demonstrate the transition driven by $\alpha$ in Fig.~\ref{fig:1}. Specifically, $\alpha$ merely replaces the coupling $K$ by $K\cos\alpha$ at linear order, and therefore it does not describe distinct steady-state behavior in identical phase oscillators for $\alpha\in(0,\pi/2)$. The emergence of $\alpha_c$ thus requires nonlinear terms beyond the linear approximation.

To gain analytical insight into the behavior that we observe in Fig.~\ref{fig:1}, 
we take the continuum limit of Eq.~(\ref{eq:model}):
\begin{equation}
    \begin{split}
            \frac{\partial \theta(r,t)}{\partial t}&=K\int d^dr' J(r-r')\Big[\\
            &\{\sin(\theta(r',t)-\theta(r,t))\}\cos\alpha\\
            &-\{\cos(\theta(r',t) - \theta(r,t)\}\sin\alpha \Big] +\eta(r,t),
    \end{split}
    \label{eq:continuum-limit-equation}
\end{equation}
where $\eta(r,t)$  is the white Gaussian noise, satisfying $\langle \eta(r,t)\eta(r',t') \rangle=D \delta^d(r-r')\delta(t-t')$. Although $K$ in Eq.~(\ref{eq:continuum-limit-equation}) can be set to unity without loss of generality, we retain it explicitly to clarify the role of the interaction.
The kernel is given by
\[
J(r-r')=\frac{((r-r')^2+r_0^2)^{-(d+\sigma)/2}}{\int d^d r' ((r-r')^2+r_0^2)^{-(d+\sigma)/2}},
\]
where finite $r_0\ne0$ regularizes the ultraviolet (UV) singularity at $r=r'$~\cite{Long-range-kernel-PRE}. 

We use the Martin-Siggia-Rose (MSR) formalism~\cite{Martin1973MSR,Janssen1976LagrangeanClassicalFieldDynamics} to obtain the following dynamic response functional
\begin{equation}
    \begin{split}
        A[\tilde{\theta},\theta]&=\int dtd^dr \ \bigg\{ \tilde{\theta} \bigg(\partial_t \theta-K\int d^dr' \ J(r-r')\Big[\\
        &\left\{\theta(r',t)-\theta(r,t)\right\}\cos\alpha \\
        &+\frac{1}{2}\left\{\theta(r',t)-\theta(r,t)\right\}^2\sin\alpha
        \Big]\bigg)-\frac{D}{2}\tilde{\theta}^2\bigg\},
    \end{split}
    \label{eq:MSR-action-real}
\end{equation}
which gives the statistical weight $P[\theta]\propto \int \mathcal{D} [i\tilde{\theta}] e^{-A[\tilde{\theta},\theta]}$, expanding the functional near $|\theta_j-\theta_i|\ll 1$ for all $\{i,j\}$ by the second order. We note that this expansion is not generally valid for long-range interacting systems. It is, however, justified for $\sigma<d$, where the contribution of spatial fluctuations scales as $\int d^d q/q^\sigma$~\cite{Defenu-Long-range-BKT}. 

By the Fourier transform, we derive the following functional~\cite{SM}
\begin{equation}
    \begin{split}
        A[\tilde{\theta},\theta]&=A_0+A_{\mathrm{int}},
    \end{split}
    \label{eq:MSR-momentum}
\end{equation}
where
\begin{equation}
\begin{split}
    A_0=&\int_q\int_\omega \bigg[\tilde{\theta}(q,\omega)\{i\omega+\kappa_0|q|^\sigma\}\theta(-q,-\omega) \\
        &  \quad \qquad -\frac{D_0}{2} \tilde{\theta}(q,\omega)\tilde{\theta}(-q,-\omega) \bigg]\\
        \label{eq:A-0}
\end{split}
\end{equation}
and
\begin{equation}
    \begin{split}
        A_{\mathrm{int}}=& \lambda_0\int_{\tilde{q},{q}}\int_{\tilde{\omega},{\omega}}\tilde{\theta}(\tilde{q},\tilde{\omega})|\tilde{q}|^\sigma \theta(q,\omega) \theta(-\tilde{q}-q, -\tilde{\omega}-\omega)\\
        & +g_0\int_{\tilde{q},{q}}\int_{\tilde{\omega},{\omega}} \tilde{\theta}(\tilde{q},\tilde{\omega})|q|^\sigma \theta(q,\omega) \theta(-\tilde{q}-q, -\tilde{\omega}-\omega),
    \end{split}
    \label{eq:A-int}
\end{equation}
where $\int_{q_1,q_2,...,q_n}=1/(2\pi)^{nd} \int d^dq_1 \int d^dq_2 ... \int d^dq_n$ and $\int_{\omega_1,\omega_2,...,\omega_n}=1/(2\pi)^{n} \int d\omega_1 \int d\omega_2 ... \int d\omega_n$ are denoted for abbreviations. 
The \textit{bare} parameters in Eqs.~(\ref{eq:MSR-momentum})--(\ref{eq:A-int}) are given by
\begingroup
\setlength{\jot}{12pt} 
\begin{subequations}
\begin{align}
\kappa_0&= c(\sigma,r_0)K\cos\alpha, \\
\lambda_0&= (c(\sigma,r_0)/2)K\sin\alpha,\\
g_0&=-c(\sigma,r_0)K\sin\alpha,
\end{align}
\label{eq:bare-parameters}
\end{subequations}
where $c(\sigma,r_0)=-\Gamma(-\sigma/2)/\left[\Gamma(\sigma/2)(2/r_0)^\sigma\right]>0$ for $\sigma\in(0,2)$~\cite{table1969-gradshteyn,Fractional-calculus-Mainardi,SM}.

The response functional derived here has the same structure as that of the Kardar--Parisi--Zhang (KPZ) equation~\cite{KPZ,Two-loop-KPZ}, except for two essential differences. First, Eqs.~(\ref{eq:A-0}) and (\ref{eq:A-int}) describe phase oscillators with long-range interactions~\cite{Joyce, Fisher-Ma-Nickel,Defenu-Long-range-BKT,Long-range-kernel-PRE,QTwisted_Lee}, whereas the short-range case was previously studied on $d$-dimensional Euclidean lattices with $d=1,2$ by mapping the corresponding differential equations to the KPZ equation~\cite{KS-KPZ-1d2d}. Second, the functional in Eq.~(\ref{eq:A-int}) contains two distinct coupling constants, $\lambda_0$ and $g_0$, in the $\tilde{\theta}\theta\theta$ vertex. This distinction arises because the factor $|\tilde{q}|^\sigma$ multiplying $\lambda_0$ scales with the auxiliary field $\tilde{\theta}(\tilde{q},\tilde{\omega})$, whereas the factor $|q|^\sigma$ multiplying $g_0$ scales with the physical field $\theta(q,\omega)$. Accordingly, $\lambda_0$ and $g_0$ must be treated separately, since the momentum scaling dimensions of the physical field $\theta$ and the auxiliary field $\tilde{\theta}$ are generally different~\cite{Martin1973MSR,Janssen1976LagrangeanClassicalFieldDynamics,tauber2014critical,Two-loop-KPZ}. As we will show below, this second feature is crucial for the renormalization group (RG) analysis for $\sigma\in(0,d)$ with $d=1$ and 2.

We perturbatively calculate $\ln \langle e^{-A_{\mathrm{int}}}\rangle$ with respect to $A_0$ by means of the cumulant expansion. This perturbative approach captures the IR behavior near $\sigma=d$ and for $\alpha \ll 1$, since the bare parameters in $A_{\mathrm{int}}[\tilde{\theta},\theta]$ [Eq.~(\ref{eq:A-int})] are proportional to $\sin\alpha$ [Eq.~(\ref{eq:bare-parameters})].

We then perform the Wilson RG~\cite{Wilson-Fisher-epsilon}, integrating out modes in the momentum shell $k\in[\Lambda/b,\Lambda]$ with $b=e^{\delta l}>1$ and the UV cutoff $\Lambda$, followed by rescaling. For $\delta l\ll 1$, this procedure yields the RG flow equations for the running couplings $D$, $\kappa$, $\lambda$, and $g$ toward the IR limit~\cite{SM}.
\begin{subequations}
\label{eq:RG-org-eqs}
\begin{equation}
    \frac{dD}{dl}=\left[\left(d+2\chi'+3z\right)+K_d\frac{g^2 D}{\kappa^3}\right]D,
    \label{eq:RG-D}
\end{equation}
\begin{equation}
    \frac{d\kappa}{dl}= \left[(z-\sigma)-2\frac{K_d}{\kappa^3}{D(g^2+\lambda^2+g\lambda)}\right]\kappa,
    \label{eq:RG-nu}
\end{equation}
\begin{equation}
    \frac{d\lambda}{dl}=\left[-\left(d+\chi'+\sigma\right) + 6K_d\frac{D}{ \kappa^3}\left\{g^2 +\lambda g +\lambda^2\right\}\right]\lambda,
    \label{eq:RG-lamdba}
\end{equation}
\begin{equation}
    \frac{dg}{dl}=\left[-\left(d+\chi'+\sigma\right)+6K_d \frac{D}{\kappa^3} \left\{ g^2 +\lambda g \right\} \right]g,
    \label{eq:RG-g}
\end{equation}
\label{eq:RG-equations-Wilson}
\end{subequations} where $K_d\equiv S_d/[2(2\pi)^d]$, with $S_d=2\pi^{d/2}/\Gamma(d/2)$ denoting the solid angle, and $\chi'$ is the scaling exponent of the physical field, defined by $\theta'=b^{\chi'}\theta$ under the rescaling $p'=bp$ and $\omega'=b^z\omega$. Here, we set $\Lambda=1$, since its value does not qualitatively affect the universal characteristics~\cite{goldenfeld2018lectures,SM}. In contrast to the KPZ equation, for which the nonlinear coupling is not renormalized due to the Galilean invariance~\cite{KPZ,Two-loop-KPZ}, both $\lambda$ and $g$ are renormalized in the present model [Eqs.~(\ref{eq:RG-org-eqs}c) and (\ref{eq:RG-org-eqs}d)]. Nevertheless, the response functional in Eq.~(\ref{eq:MSR-action-real}), derived from the dynamics governed by Eq.~(\ref{eq:continuum-limit-equation}), is invariant under the global shift $\theta\to\theta+c$, where $c$ is an arbitrary constant. The corresponding Ward identity ensures that the mass term $\tilde{\theta}\theta$ is not generated at any order in perturbation theory~\cite{Shift-gauged-symmetry-KPZ}.

Introducing the dimensionless couplings
\begin{equation}
\tilde{g}_1=\frac{g^2D}{\kappa^3}, \qquad
\tilde{g}_2=\frac{\lambda g D}{\kappa^3}, \qquad
\tilde{g}_3=\frac{\lambda^2 D}{\kappa^3},
\label{eq:dimensionless-couplings}
\end{equation}
the RG equations become~\cite{SM}
\begin{subequations}
    \begin{equation}
        \frac{d\tilde{g}_1}{dl}
=\tilde{g}_1\left[(-d+\sigma)+K_d\left\{19\tilde{g}_1 + 18\tilde{g}_2 +6 \tilde{g}_3\right\} \right],
        \label{eq:g_1}
    \end{equation}
    \begin{equation}
       \frac{d\tilde{g}_2}{dl}
=\tilde{g}_2\left[(-d+\sigma)+K_d\left\{19\tilde{g}_1 + 18\tilde{g}_2+12\tilde{g}_3\right\} \right],
        \label{eq:g_2}
    \end{equation}
    \begin{equation}
        \frac{d\tilde{g}_3}{dl}
=\tilde{g}_3\left[(-d+\sigma)+K_d\left\{19\tilde{g}_1 +18\tilde{g}_2+18\tilde{g}_3\right\} \right].
        \label{eq:g_3}
    \end{equation}
    \label{eq:RG-eqs}
\end{subequations}

First, Eqs.~(\ref{eq:g_1})--(\ref{eq:g_3}) have the trivial fixed point
\begin{equation}
    \left(\tilde{g}_1,\tilde{g}_2,\tilde{g}_3\right)=\left(0,0,0\right),
    \label{eq:trivial}
\end{equation}
whose eigenvalues of the linear stability matrix are degenerated to $(-d+\sigma)$. This fixed point is therefore stable for $\sigma<\sigma_c=d$. This result is consistent with the region of synchronization in the phase diagram numerically obtained in Fig.~\ref{fig:1}. Specifically, at the intersection of the physical line defined by the bare parameters [Eqs.~(\ref{eq:bare-parameters})] with the \textit{fixed} point, all dimensionless couplings [Eq.~(\ref{eq:dimensionless-couplings})] are proportional to the following \textit{effective} coupling constants:
\[
    \frac{D_0}{K}\tan^2\alpha\sec\alpha.
    \label{eq:propto-trivial-fixed}
\]
That is, at the trivial IR fixed point [Eq.~(\ref{eq:trivial})] all nonreciprocal couplings flow to zero, so the IR state is that of the equilibrium model at $\alpha=0$ with finite $K$ and $D_0$, for which the Mermin--Wagner theorem and its generalization by Cassi place the threshold at $\sigma_c=d$ for $d=1$ and $2$, i.e., $d_s=2$~\cite{Mermin-Wagner,Mermin-classical,Cassi-MerminWagner} (see End Matter). Conversely, in $\sigma<d$ ($d_s>2$) for $d=1$ and 2, spontaneous synchronization is guaranteed at sufficiently low but finite $T=D_0/2$~\cite{Inverse-Mermin-Wagner-Burioni}. Thus, nonreciprocity is irrelevant in the synchronized phase shown in Fig.~\ref{fig:1}.

Nontrivial fixed points are given by
\begin{subequations}
    \begin{equation}
    \left(\tilde{g}_1,\tilde{g}_2,\tilde{g}_3\right)=\left(0,0,\frac{1}{18K_d}(d-\sigma)\right),
        \label{eq:nontrivial-g3}
    \end{equation}
    \begin{equation} \left(\tilde{g}_1,\tilde{g}_2,\tilde{g}_3\right)=\left(\tilde{g}_1^*,\frac{1}{18K_d}\left\{(d-\sigma)-19K_d\tilde{g}_1^*\right\},0\right).
    \label{eq:nontrivial-g1-2}
    \end{equation}
\end{subequations}
We now identify the \textit{physically admissible} fixed points of Eqs.~(\ref{eq:nontrivial-g3}) and (\ref{eq:nontrivial-g1-2}) from the definitions of the dimensionless couplings [Eq.~(\ref{eq:dimensionless-couplings})]. First, both $\tilde{g}_1$ and $\tilde{g}_3$ must be nonnegative. Second, since $\tilde{g}_2=\lambda g D/\kappa^3$, it must vanish whenever either $\tilde{g}_1=0$ with $\tilde{g}_3\ne0$ or $\tilde{g}_1\ne0$ with $\tilde{g}_3=0$. The physically accessible nontrivial fixed point is therefore either Eq.~(\ref{eq:nontrivial-g3}) or
\begin{equation}
   \left(\tilde{g}_1,\tilde{g}_2,\tilde{g}_3\right)=\left(\frac{1}{19K_d}(d-\sigma),0,0\right).
    \label{eq:physically-determined-ntrfixedpoint}
\end{equation}
The first nontrivial fixed point, Eq.~(\ref{eq:nontrivial-g3}), is unstable against perturbations along the $\tilde{g}_3$ direction for $\sigma < d$, but stable along the other two eigendirections [Fig.~\ref{fig:2}]. In contrast, the second fixed point, Eq.~(\ref{eq:physically-determined-ntrfixedpoint}), is unstable along the $\tilde{g}_1$ axis and linearly marginal along the other two eigendirections. Using center-manifold theory, however, we confirm that this second fixed point is in fact unstable in all three eigendirections~\cite{SM}. Thus, the critical manifold is determined solely by Eq.~(\ref{eq:nontrivial-g3}), as schematically shown in Fig.~\ref{fig:2}.

\begin{figure}[t]
    \centering
    \includegraphics[width=0.5\linewidth]{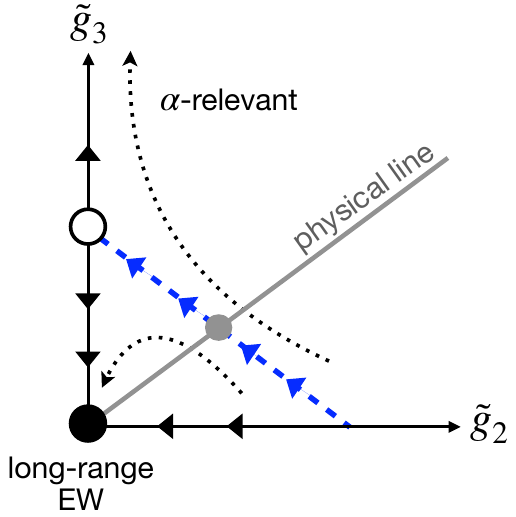} \captionsetup{justification=raggedright,singlelinecheck=false}
    \caption{Schematic RG flow in coupling-constant space for $\sigma\in(0,d)$ with $d=1$ and 2. As discussed in the main text, the $\tilde{g}_1$ direction does not affect the critical behavior; we therefore show the $(\tilde{g}_2,\tilde{g}_3)$ plane. 
    Arrows indicate RG flow directions. The filled circle at the origin denotes the trivial stable fixed point. The open circle on the $\tilde{g}_3$ axis denotes the nontrivial fixed point [Eq.~(\ref{eq:nontrivial-g3})], around which the critical manifold is locally defined and shown as the blue dashed line. This manifold intersects the physical line of the present system [Eqs.~(\ref{eq:bare-parameters})], shown as the gray straight line through the origin. Dotted arrows indicate schematic RG trajectories. 
    }
    \label{fig:2}
\end{figure}

We obtain an approximated closed-form expression of $\alpha_c$ from the intersection of the critical manifold with the ``physical line'' of our bare parameters defined by Eqs.~(\ref{eq:bare-parameters}) (see Fig.~\ref{fig:2}). Although the critical manifold derived from the linearized flow does not provide global information about the RG trajectories, it is sufficient to describe the phase transition for $\alpha\ll1$ near $\epsilon=d-\sigma=0$. Finding this intersection, we obtain~\cite{SM}
\begin{equation}
    |\alpha_c|\approx \sqrt{G\left[\frac{-\Gamma(-\sigma/2)}{\Gamma(\sigma/2)}\left(d-\sigma\right)\right]},
    \label{eq:alpha_c-text}
\end{equation}
where
\begin{equation}
    G\equiv  \frac{K}{D_0}\frac{20(r_0/2)^\sigma}{183K_d\Lambda^{d-\sigma}}
    \label{eq:G-coeff-text}
\end{equation}
(see the insets of Figs.~\ref{fig:1}(a) and \ref{fig:1}(c)).

\textit{Criticality at IR fixed points--} We describe the critical exponents of Eq.~(\ref{eq:model}) for $\sigma\in(0,d)$ with $d=1$ and 2. The synchronized regime $\alpha<\alpha_c$, which we analytically show to be equivalent to $\alpha=0$ in the IR limit, flows toward the long-range EW fixed point: from Eqs.~(\ref{eq:RG-equations-Wilson}), we obtain the dynamical exponent $z=\sigma$ and the roughness exponent $\chi=-(\chi'+d+z)=(\sigma-d)/2$~\cite{long-range-EW} at the trivial fixed point. This also holds at the level of the dynamical equation; in the regime $|\theta_j-\theta_i|\ll 1 $ for all $\{i,j\}$, Eq.~(\ref{eq:continuum-limit-equation}) with $\alpha=0$ reduces to 
\[
    \frac{\partial \theta(r,t)}{\partial t}=K\int d^dr' J(r-r')(\theta(r',t)-\theta(r,t))+\eta(r,t),
\]
which is the long-range EW equation~\cite{long-range-EW}.

By contrast, at $\alpha=\alpha_c$, corresponding to the nontrivial fixed point, Eqs.~(\ref{eq:RG-equations-Wilson}) yield 
\begingroup
\setlength{\jot}{12pt} 
\begin{subequations}
\begin{align}
 z&=\sigma+{(d-\sigma)}/{9}, \\
\chi&={(\sigma-d)}/{2}+{(d-\sigma)}/{18}, \\
1/\nu &= (d-\sigma),
\end{align}
\label{eq:analytical-critical-exponents}
\end{subequations}
where $\nu$ is the correlation length exponent~\cite{goldenfeld2018lectures, Two-loop-KPZ}. 
The critical exponents in Eqs.~(\ref{eq:analytical-critical-exponents}), which exhibit nonzero one-loop corrections of $O(\epsilon)$ for $\sigma<d$, demonstrate that the critical behavior near $\alpha=\alpha_c$ belongs to a universality class distinct from that of the long-range EW fixed point. 

We further perform finite-size scaling analysis to numerically corroborate the critical exponents in Eqs.~(\ref{eq:analytical-critical-exponents}), with representative data collapses shown in Fig.~\ref{fig:3}; see End Matter for details.

\begin{figure}[t]
     \centering
     \includegraphics[width=0.85\linewidth]{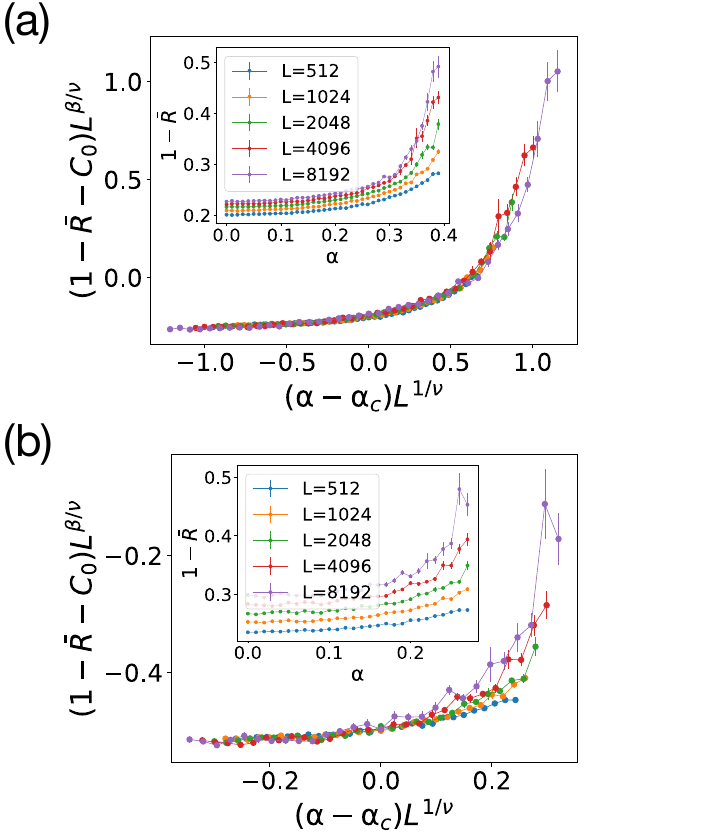}
     \captionsetup{justification=raggedright,singlelinecheck=false}
     \caption{Representative data collapses on the synchronized side ($\alpha<\alpha_c$) for $d=1$, obtained from finite-size scaling analysis using the linear system size $L$ (see End Matter). The data collapses are obtained using the critical exponents in Eqs.~(\ref{eq:analytical-critical-exponents}) and visual fits with $\alpha_c=0.20$ and $C_0=0.28$ for $\sigma=0.8$ in (a), and with $\alpha_c=0.14$ and $C_0=0.53$ for $\sigma=0.9$ in (b). Each panel and its inset share the same legend. The insets show $1-\bar{R}$ obtained by numerical integration with $t_{\mathrm{avg}}=2\times10^4$ and $t_r=1.8\times10^5$, starting from random initial phase configurations. We set $D=0.36$ and $r_0=0.2$. The symbol and error bar represent the mean and standard error over 15 noise and initial-phase realizations.}
     \label{fig:3}
\end{figure}

\textit{Discussion--} We have shown that phase transitions in the noisy Kuramoto--Sakaguchi model with identical oscillators for $|\alpha|\in[0,\pi/2)$ are governed by the exponent $\sigma$, which controls the power-law decay of the long-range interactions. Our results provide a theoretical framework for analytically understanding critical phenomena induced by nonreciprocity in long-range interacting systems. A natural direction for future work is to determine whether the response functional in Eqs.~(\ref{eq:MSR-momentum})--(\ref{eq:A-int}) defines a new universality class by evaluating the scaling exponents with higher precision. Furthermore, extending our analysis to other nonreciprocal oscillator models would establish the robustness of our results for systems with long-range interactions~\cite{Future-work}.

Moreover, we numerically observe qualitatively simliar behaivor in phase diagram of complex networks~\cite{LRRR} (see Ref.~\cite{SM}). This finding suggests that our theory for long-range connections may extend beyond translationally invariant lattices to more general geometries.

\medskip
\begin{acknowledgments}
\textit{Acknowledgments}---This study was supported by JST SPRING, Grant No. JPMJSP2108.

\end{acknowledgments}

\bibliography{apssamp}

\clearpage
\onecolumngrid
\begin{center}
  {\large\textbf{End Matter}}\\[3pt]
\end{center}
\vspace{8pt}
\twocolumngrid

\appendix
\setcounter{equation}{0}
\renewcommand{\theequation}{E\arabic{equation}}
\textit{Response and correlation propagators--} The Gaussian part of $A[\tilde{\theta},\theta]$, i.e., $A_0$, is 
\begin{equation}
    A_0[\tilde{\theta},\theta]=\frac{1}{2} \int_k \int_\omega \big(\tilde{\theta}(q,\omega), \theta(q,\omega)\big)\mathbf{A}(q,\omega)\begin{pmatrix}\tilde{\theta}(-q,-\omega)\\
\theta(-q,-\omega)\end{pmatrix},
    \label{eq:Gaussian-part}
\end{equation}
where 
\begin{equation}
\mathbf{A}(q,\omega)= \begin{pmatrix} -D_0 & i\omega +\kappa_0 |q|^\sigma \\
-i\omega +\kappa_0|q|^\sigma & 0
\end{pmatrix}.
    \label{eq:Gaussian-matrix}
\end{equation}
Then the inverse matrix of $A$ is 
\begin{equation}
\mathbf{A}^{-1}(q,\omega)= 
\begin{pmatrix} 0 & \frac{1}{-i\omega+\kappa_0|q|^\sigma} \\
\frac{1}{i\omega+\kappa_0|q|^\sigma} & D_0\left|\frac{1}{i\omega+\kappa_0|q|^\sigma}\right|^2
\end{pmatrix},
    \label{eq:S-A-Gaussian2}
\end{equation}
from which we obtain the response and correlation propagators as follows:
\begin{subequations}
   \label{eq:propagators}
    \begin{equation}
        G_0(q,\omega)\equiv\langle \tilde{\theta}(q,\omega)\theta(-q,-\omega)\rangle =\frac{1}{-i\omega + \kappa_0|q|^\sigma},
    \label{eq:G0}
    \end{equation}
    \begin{equation}
       C_0(q,\omega)\equiv\langle \theta(q,\omega)\theta(-q,-\omega)\rangle=D_0|G_0(q,\omega)|^2.
    \label{eq:C0}
    \end{equation}
\end{subequations}

The response propagator in $t$-domain is
\begin{equation}
        G_0(q,t) = \frac{1}{2\pi}\int_{-\infty}^{\infty} d\omega \frac{e^{-i\omega t}}{-i\omega+\kappa_0 |q|^\sigma}=\exp[-\kappa_0|q|^\sigma t]u(t),
\end{equation}
where $u(t)$ is the Heaviside step function.
Since $\kappa_0>0$ for $\alpha\in[0,\pi/2)$, $G_0$ provides a stable description of the dynamical renormalization-group flow for $t>0$. 
\medskip

\textit{Thermodynamic-limit value of $\sigma_c$ for $d=1$ and $d=2$--} Here, we demonstrate that the thermodynamic limit of the critical value of $\sigma$ for $d=1$ and $d=2$ is $\sigma_c=d$, from the Mermin-Wagner theorem~\cite{Mermin-classical,Cassi-MerminWagner}. The theorem describes that the continuous symmetry of quantum Heisenberg ferromagnetic models cannot be spontaneously broken under any nonzero temperature in the lattice whose dimension is less than or equal to two~\cite{Mermin-Wagner}, and Mermin found that the dependence on the dimension is the same for classical $O(n)$ models~\cite{Mermin-classical}.
Beyond translationally invariant lattices, Cassi rigorously showed that the same lower critical dimension is governed by the spectral dimension $d_s$: spontaneous magnetization is absent on general graphs with spectral dimension $d_s\le 2$ in the thermodynamic limit~\cite{Cassi-MerminWagner}.

The spectral dimension $d_s$ is defined by the return probability of a random walker~\cite{Rammal1984RandomWalkFractals}:
\begin{equation}
    P_0(t)\sim t^{-d_s/2}, \quad t \gg 1,
    \label{eq:P0(t)}
\end{equation}
or the cumulative distribution of the eigenvalues~\cite{Universal-spectral-dimension}:
\begin{equation}
    \rho_c(\lambda)\simeq \lambda^{d_s/2}, \quad \lambda\ll 1.
    \label{eq:rho(lambda)}
\end{equation}
of the normalized Laplacian $\mathbf{L}$ whose elements are given by
\begin{equation}
    L_{ij}=\delta_{ij}-\frac{J_{ij}}{k_i},
    \label{eq:normalized-L}
\end{equation}
as a $N\times N$ matrix, where $J_{ij}$ is the (weighted) adjacency matrix, and $k_i$ is the degree of node $i$.

For translationally invariant lattices, the Laplacian operator is written by
\begin{equation}
(L\phi)(\mathbf{x})=\sum_{\mathbf{y}}J(\mathbf{x}-\mathbf{y})[\phi(\mathbf{x})-\phi(\mathbf{y})],
\label{L-operator}
\end{equation}
and its eigenvalue $\lambda(\mathbf{p})$ is given by
\begin{equation}
\lambda(\mathbf{p})=\frac{1}{k}\sum_{\mathbf{r}}J(\mathbf{r})[1-e^{-i\mathbf{p}\cdot \mathbf{r}}],
\end{equation}
where $k\equiv \sum_\mathbf{r}J(\mathbf{r})$, with its eigenvector
\begin{equation}
   \phi_\mathbf{p}(\mathbf{x})=e^{-i\mathbf{p}\cdot \mathbf{x}} .
\end{equation}

To obtain $d_s$ of the fully connected weighted graph with $J(r-r')$ of the main text as a function of $\sigma$ in closed-form expression, we apply the continuum limit. Specifically, the eigenvalue $\lambda$ becomes
\[
\begin{split}
    \lambda(\mathbf{p})&=\frac{1}{\int d^dr' (r'^2+r_0^2)^{-(d+\sigma)/2}}\int d^dr' \ \frac{ 1-e^{-i\mathbf{p}\cdot \mathbf{r'} }}{(r'^2+r_0^2)^{(d+\sigma)/2}}\\
    &= J(0)-J(\mathbf{p}),
\end{split}
\]
where we introduce the regularization by finite $r_0$, as described in the main text. We note that the IR scaling of $\lambda(\mathbf{p})$ remains proportional to $|\mathbf{p}|^\sigma$ even for $r_0\ne0$ (see Ref.~\cite{SM}).

We can then derive the spectral dimension $d_s$ from the analytical form of the density of states (DOS) for the eigenvalue $\lambda(\mathbf{p})$. The integrated density of states satisfies
\begin{equation}
    \rho_c(\lambda)
    \propto \int_{|p'|\le p(\lambda)} \frac{d^dp'}{(2\pi)^d}
    \propto [p(\lambda)]^d
    \propto \lambda^{d/\sigma},
\end{equation}
since $\lambda\propto |\mathbf{p}|^\sigma$. 
Therefore, from Eq.~(\ref{eq:rho(lambda)}), we obtain 
\begin{equation}
d_s =
\begin{cases}
2d/\sigma, & \mathrm{if\ } 0 < \sigma < 2,\\
d, & \mathrm{if\ } \sigma \ge 2.
\end{cases}
    \label{eq:d_s_annealed}
\end{equation}
For $\sigma\ge 2$, the spectral dimension is $d_s=d$, since the $|\mathbf{p}|^\sigma$ term is no longer the leading contribution to the IR expansion~\cite{Joyce, Fisher-Ma-Nickel}.

With Eq.~(\ref{eq:d_s_annealed}), for $d=1$ and $d=2$, the critical value of $\sigma$ is $\sigma_c=d$, corresponding to $d_s=2$ in the thermodynamic limit~\cite{Mermin-Wagner,Mermin-classical,Cassi-MerminWagner,Inverse-Mermin-Wagner-Burioni}; we note that the criterion $\sigma_c=d$ applies only when the embedding geometry of the long-range weighted edges is translationally invariant and either one- or two-dimensional ($d=1$ or $2$), since $\sigma_c=d$ implies $\sigma_c>2$ for $d>2$; the corresponding threshold is therefore physically absent in the IR regime for $d>2$. Thus, at $\alpha=0$, a synchronization transition driven by varying $\sigma$ occurs only for $d=1$ and $d=2$.
\medskip

\textit{Criterion of the nonzero noise strength--} 
To establish a criterion relating the noise strength $D$, the linear system size $L$, and the spatial dimension $d$, we note that, at $\alpha=0$, the spatial phase fluctuation $\langle\theta^2\rangle$ scales as
\begin{equation}
    \langle\theta^2\rangle \propto \frac{D}{K}\int_{2\pi/L}^{2\pi/r_0}
    \frac{d^d p}{(2\pi)^d}\frac{1}{|\mathbf{p}|^\sigma}
    \sim \frac{D}{K}L^{\sigma-d},
\end{equation}
which diverges for $\sigma>d$ in the thermodynamic limit at nonzero $D$ and finite $K$~\cite{Defenu-Long-range-BKT}. For finite $L$, however, the spatial phase fluctuation remains bounded and may be insufficiently large to destroy synchronization at a given $D$. Consequently, synchronization can persist even for $\sigma>d$. Thus, although a small value of $D$ is required for synchronization to emerge for $\sigma<d$, $D$ must also be chosen such that the spatial phase fluctuation becomes sufficiently large for $\sigma>d$ at the system size considered. Because this finite-size effect is controlled by the linear system size $L$ rather than the number of nodes $N$, we choose $D$ accordingly to reveal the critical behavior in finite-size numerical simulations.

\medskip

\textit{Finite-size scaling ansatz--} 
  For $d=1$, our DRG analysis is not associated with the crossover to short-range criticality because the perturbative expansion is performed near $\sigma=d=1$. By contrast, for $d=2$, the criticality near a threshold $\sigma^* \simeq 2$ may be strongly affected by crossover effects~\cite{Fisher-Ma-Nickel,Sak-long-range-crossover}. In this regard, for $d=1$, we numerically support the critical exponents with $O(\epsilon)$ corrections in Eqs.~(\ref{eq:analytical-critical-exponents}) using finite-size scaling analysis. Specifically, because our perturbative DRG analysis assumes the regime $|\theta_j-\theta_i|\ll 1$, the theory developed in this Letter describes the power-law behavior upon approaching criticality from the synchronized side. This is consistent with the negative value of $\chi$ in Eq.~(\ref{eq:analytical-critical-exponents}b) for $\sigma<d$, which renders $\chi$ subleading to the zeroth-order contribution. We therefore employ the following scaling ansatz for the order parameter below the critical point, $\alpha<\alpha_c$, to obtain data collapes with the critical exponents from the numerical results in the long-time regime:
\begin{equation}
1-R (\alpha)=C_0+L^{-\beta/\nu}F\left(|\alpha_c-\alpha| L^{1/\nu}\right),
\label{eq:finite-size scaling-conventional-form}
\end{equation}
where $C_0\equiv\lim_{L\to\infty}[1-R(\alpha_c)]$ denotes a nonuniversal background contribution, and $L$ is the linear size of the lattice~\cite{goldenfeld2018lectures}. Here, $\beta$ is the critical exponent defined by $|1-R-C_0|\sim|\alpha_c-\alpha|^{\beta}$ near criticality. We see that $-\beta/\nu$ approxes $2\chi$ for $d=1$ near the synchronization, as $R\approx 1-W^2/2$, where $W^2\equiv\langle(\theta_j-\langle\theta\rangle)^2\rangle$ is the roughness and $\langle\cdot\rangle$ denotes a spatial average~\cite{katoh-kori-W2-sync}. We restrict our numerical analysis to the static critical exponents, as a reliable estimate of the dynamic exponent requires appropriate cutoffs at both short and long times and is therefore considerably less controlled than the steady-state analysis~\cite{Marinari-Enzo-Paraisi-critical}.

\clearpage

\setcounter{page}{1}
\renewcommand{\thepage}{S\arabic{page}}
\onecolumngrid 

\setcounter{equation}{0}
\setcounter{figure}{0}
\setcounter{table}{0}

\renewcommand{\theequation}{S\arabic{equation}}
\renewcommand{\thefigure}{S\arabic{figure}}  
\renewcommand{\thetable}{S\arabic{table}}     

\begingroup
\renewcommand{\thefootnote}{}%
\footnotetext{* Contact author: minwoo-bae@g.ecc.u-tokyo.ac.jp}%
\endgroup

\begin{center}
    \textbf{\large Supplemental Material for\\
``Criticality of nonreciprocal phase oscillators with long-range interactions''}\\[0.5cm]

Minwoo Bae$^{1,\ *}$ and Hiroshi Kori$^{1}$\\[0.2cm]

\textit{Department of Complexity Science and Engineering, Graduate School of Frontier Sciences, the University of Tokyo, Chiba 277-8561, Japan}\\[0.2cm]

(Dated: \today)
\end{center}

\maketitle

\section{A. The kernel $J(p)$ in the momentum space}
We obtain the formula of the long-range kernel in the momentum space, i.e., $J(p)$, by the Fourier transform with $J(r)=(r^2+r_0^2)^{-(d+\sigma)/2}/\int d^dr' ((r')^2+r_0^2)^{-(d+\sigma)/2}$.
\begin{equation}
    \begin{split}
    J(p)&=\frac{1}{\int d^dr' ((r')^2+r_0^2)^{-(d+\sigma)/2}}\int d^dr\ e^{-i\mathbf{p}\cdot\mathbf{r} } (r^2+r_0^2)^{-(d+\sigma)/2}\\
    &=\frac{1}{J_0} \int_0^\infty dr \ \frac{r^{d-1}}{(r^2+r_0^2)^{(d+\sigma)/2}} \int d\Omega_{d-1} \ e^{-i|\mathbf{p}|r\cos\theta},
    \end{split}
    \label{eq:radial-and-angular-Jp}
\end{equation}
where we denote by $\int d\Omega_{d-1}$ the angular integral over the $(d-1)$-dimensional unit sphere transverse to the radial direction of $\boldsymbol{r}$, and define $J_0\equiv \pi^{d/2}\Gamma(\sigma/2)/[r_0^\sigma\Gamma((d+\sigma)/2)]$.
The angular integral term $A_d(pr)\equiv\int d\Omega_{d-1}\ e^{-i|\mathbf{p}|r\cos\theta}$ is
\begin{equation}
    \begin{split}
    A_d(pr)&=S_{d-1}\int_0^\pi d\theta \ (\sin\theta)^{d-2}e^{-i|\mathbf{p}|r\cos\theta}\\
    &=\frac{2\pi^{(d-1)/2}}{\Gamma\left(\frac{d-1}{2}\right)}\int_{-1}^1 dt \ (1-t^2)^{(d-3)/2}e^{-i|\mathbf{p}|rt}\\
    &=(2\pi)^{d/2}(|\mathbf{p}|r)^{1-d/2}J_{d/2-1}(|\mathbf{p}|r),
    \end{split}
    \label{eq:angular-Jp}
\end{equation}
where $S_d$ is the solid angle. In the last equality, we use the integral representation of the Bessel functions of the first kind $J_\kappa(z)$, given by formula 8.411(10) of Ref.~\cite{table1969-gradshteyn}. Eqs.~(\ref{eq:radial-and-angular-Jp}) and (\ref{eq:angular-Jp}) yield
\begin{equation}
    J(p)=\frac{(2\pi)^{d/2}}{J_0}|\mathbf{p}|^{1-d/2}\int_0^\infty dr \ r^{d/2}(r^2+r_0^2)^{-(d+\sigma)/2}J_{d/2-1}(|\mathbf{p}|r).
    \label{eq:Jp-before-final}
\end{equation}
Using formula 6.565(4) of Ref.~\cite{table1969-gradshteyn}, Eq.~(\ref{eq:Jp-before-final}) becomes
\begin{equation}
    J(p)=\frac{2\pi^{d/2}|\mathbf{p}|^{\sigma/2}}{J_0(2r_0)^{\sigma/2}\Gamma\left(\frac{\sigma+d}{2}\right)}K_{\sigma/2}(|\mathbf{p}|r_0),
\end{equation}
where $K_\kappa(xz)$ is the modified Bessel functions of the second kind. It is also known that $K_\kappa(z)$ is approximated by the following two leading order terms:
\begin{equation}
    K_\kappa (z) \approx \frac{1}{2}
    \Gamma(\kappa)\left(\frac{z}{2}\right)^{-\kappa}+\frac{1}{2}\Gamma(-\kappa)\left(\frac{z}{2}\right)^\kappa,
\end{equation}
for $z\ll 1$ where $\kappa\in(0,1)$~\cite{Fractional-calculus-Mainardi}. Here, $\kappa\in (0,1)$ corresponds to $\sigma\in(0,2)$, where long-range weights are relevant~\cite{Fisher-Ma-Nickel}.
We thus obtain the kernel $J(p)$ to describe a relevant behavior in the IR regime:
\begin{equation}
    J(p)\approx 1-c(\sigma,r_0)|\mathbf{p}|^\sigma,
    \label{eq:p-dependence-J(p)}
\end{equation}
where
\begin{equation}
    c(\sigma,r_0)=\frac{-\Gamma\left(-\frac{\sigma}{2}\right)}{\Gamma\left(\frac{\sigma}{2}\right)(2/r_0)^{\sigma}}.
    \label{eq:c-sigma-def}
\end{equation}
Note that $c(\sigma,r_0)$ is positive for $\sigma\in(0,2)$.

\section{B. Derivation of Equations~(\ref{eq:MSR-momentum})--(\ref{eq:propagators})}

We rewrite Eq.~(\ref{eq:MSR-action-real}) of the main text for convenience:
\begin{equation}
    \begin{split}
        A[\tilde{\theta},\theta]&=\int dtd^dr \ \bigg\{ \tilde{\theta} \bigg(\partial_t \theta-K\int d^dr' \ J(r-r')\Big[\\
        &\left\{\theta(r',t)-\theta(r,t)\right\}\cos\alpha \\
        &+\frac{1}{2}\left\{\theta(r',t)-\theta(r,t)\right\}^2\sin\alpha
        \Big]\bigg)-\frac{D}{2}\tilde{\theta}^2\bigg\},
    \end{split}
\end{equation}
which gives the statistical weight $P[\theta]\propto \int \mathcal{D} [i\tilde{\theta}] e^{-A[\tilde{\theta},\theta]}$, related to the generating functional
\begin{equation}
\begin{split}
    Z[j,j^*]&=\iint \mathcal{D}[i\tilde{\theta}]\mathcal{D}[\theta]\exp\bigg\{-A[\tilde{\theta},\theta]\\
    &+\iint dt d^dr \left[j^*(r,t)\tilde{\theta}(r,t)+j(r,t)\theta(r,t)\right] \bigg\}.
    \end{split}
    \label{eq:Z-jj}
\end{equation}

Then, we obtain the momentum representation of the response functional $A[\tilde{\theta}, \theta]$ by the Fourier transform.
For the first term, we obtain $(2\pi)^{-d-1}\iint d^dq d\omega \ \tilde{\theta}(q,\omega)(i\omega)\theta(-q,-\omega)$. The second term is obtained by
\begin{equation}
    \begin{split}
        &-\iint dtd^dr \ K\tilde{\theta}(r,t)\int d^dr' \ J(r-r')\cos\alpha\left[\theta(r',t)-\theta(r,t)\right]\\
        &=\iint dtd^dr  \ \frac{(-K\cos\alpha)}{(2\pi)^{3d}}  \bigg[ \int d^dr' \int d^d \tilde{q} \ e^{i\tilde{q}r}\tilde{\theta}(\tilde{q},t)\int d^dq \ e^{iq(r-r')}J(q)  \int d^dq' e^{iq'r'}\theta(q',t)\\ 
        &- \int d^d \tilde{q} \ e^{i\tilde{q}r}\tilde{\theta}(\tilde{q},t)\int  d^dq' e^{iq'r} \theta(q',t) J(q=0)\bigg]\\
        &=\iint dtd^dr  \ \frac{(-K\cos\alpha)}{(2\pi)^{3d} } \bigg[ \int d^dr'   \iiint d^d\tilde{q} d^dq' d^dq\ e^{ir'(q'-q)}e^{ir(\tilde{q}+q)}J(q)\tilde{\theta}(\tilde{q},t) \theta(q',t)\\
        & \quad -J(0)\iint d^d\tilde{q} d^dq' \tilde{\theta}(\tilde{q},t)\theta(q',t)e^{ir(\tilde{q}+q')} \bigg]\\
        &=\iint dt  \ \frac{(-K\cos\alpha)}{(2\pi)^d } \bigg[\int d^dq'\ \tilde{\theta}(-q',t)J(q')\theta(q',t)-J(0)\int d^dq'\ \tilde{\theta}(-q',t) \theta(q',t) \bigg]\\
        &=\iint dt  \ \frac{(-K\cos\alpha)}{(2\pi)^d } \left[\int d^dq\ \tilde{\theta}(-q,t)\left\{J(q)-J(0)\right\}\theta(q,t)\right] \\
        &=\frac{1}{(2\pi)^d} \int dt \left[\int d^dq \ \tilde{\theta}(-q,t)(\kappa_0|q|^\sigma)\theta(q,t)\right],
    \end{split}
\end{equation}
where $\kappa_0=c(\sigma,r_0)K\cos\alpha$.

To further apply the Fourier transform about $t$, the result is equal to $(2\pi)^{-d-1}\iint d^dqd\omega\ \tilde{\theta}(q,\omega)(\kappa_0|q|^\sigma)\theta(-q,-\omega)$.

The third term is 
\begin{equation}
    \begin{split}
        &-\int d^dr\ \tilde{\theta}(r)\frac{K\sin\alpha}{2}\int d^dr' \ J(r-r')\Big[\theta(r')-\theta(r)\Big]^2\\
        &=\int d^dr\ \tilde{\theta}(r)\frac{(-K\sin\alpha)}{2}\int d^dr' \ J(r-r')\left[\theta(r')^2-2\theta(r')\theta(r) +\theta(r)^2\right],
\end{split}
\label{eq:third-term}
\end{equation}
for which we suggest each result of respective three terms and incorporate for the entire response functional as follows.

The first one is given by
\begin{equation}
    \begin{split}
&\int d^dr \ \tilde{\theta}(r)\frac{(-K\sin\alpha)}{2}\int d^dr' \ J(r-r')\theta(r')^2\\
&=\frac{(-K\sin\alpha)}{2(2\pi)^{4d}}\int d^dr \ \int d^d\tilde{q}\ e^{i\tilde{q}r}\tilde{\theta}(\tilde{q},t) \int d^dr' \iiint d^dq d^dq' d^dq'' e^{iq(r-r')}J(q) e^{iq'r'}\theta(q',t)e^{iq'' r'}\theta(q'',t)\\
&= \frac{(-K\sin\alpha)}{2(2\pi)^{4d}} \iint d^dr d^dr' \int d^d\tilde{q} \iiint d^dq d^dq' d^dq''\ \tilde{\theta}(\tilde{q},t) J(q) \theta(q',t) \theta(q'',t) e^{ir(\tilde{q}+q)}e^{ir'(q''+q'-q)}
\\
&= \frac{(-K\sin\alpha)}{2(2\pi)^{3d}}\int d^dr' \iiint d^dq d^dq' d^dq'' \tilde{\theta}(-q,t)J(q)\theta(q',t)\theta(q'',t)e^{ir'(q''+q'-q)}\\
&=\frac{(-K\sin\alpha)}{2(2\pi)^{2d}} \iint d^dq d^dq'  \tilde{\theta}(-q,t)J(q)\theta(q',t)\theta(q-q',t) \\
&=\frac{(-K\sin\alpha)}{2(2\pi)^{2d}} \iint d^d\tilde{q} d^dq\  \tilde{\theta}(\tilde{q},t)J(-\tilde{q})\theta(q,t)\theta(-\tilde{q}-q,t). 
    \end{split}
\end{equation}
The second one is given by
\begin{equation}
    \begin{split}
&\int d^dr \ \tilde{\theta}(r){K\sin\alpha}\int d^dr' \ J(r-r')\theta(r')\theta(r)\\
&=\frac{K\sin\alpha}{(2\pi)^{4d}}\int d^dr \int d^d\tilde{q} \ e^{i\tilde{q}r}\tilde{\theta}(\tilde{q},t) \int d^dr' \iiint d^dq d^dq' d^dq'' e^{iq(r-r')} J(q) e^{iq'r'}\theta(q',t)e^{iq''r}\theta(q'',t)\\
&=\frac{K\sin\alpha}{(2\pi)^{4d}}\iint d^dr d^dr' \int d^d\tilde{q}\iiint d^dqd^dq'd^dq'' \tilde{\theta}(\tilde{q},t)J(q)\theta(q',t)\theta(q'',t)e^{ir(\tilde{q}+q+q'')}e^{ir'(q'-q)}\\
&=\frac{K\sin\alpha}{(2\pi)^{3d}} \int d^dr  \int d^d\tilde{q}\ \iint d^dq' d^dq'' \tilde{\theta}(\tilde{q},t)J(q')\theta(q',t)\theta(q'',t)e^{ir(\tilde{q}+q'+q'')} \\
&=\frac{K\sin\alpha}{(2\pi)^{2d}}   \iint d^d\tilde{q}d^dq  \tilde{\theta}(\tilde{q},t)J(q)\theta(q,t)\theta(-\tilde{q}-q,t).
    \end{split}
\end{equation}
The last one is given by
\begin{equation}
    \begin{split}
&\int d^dr \ \tilde{\theta}(r)\frac{(-K\sin\alpha)}{2}\int d^dr' \ J(r-r')\theta(r)^2\\
&=\frac{(-K\sin\alpha)}{2(2\pi)^{4d}}\int d^dr \ \int d^d\tilde{q}\ e^{i\tilde{q}r}\tilde{\theta}(\tilde{q},t) \int d^dr' \iiint d^dq d^dq' d^dq'' e^{iq(r-r')}J(q) e^{iq'r}\theta(q',t)e^{iq'' r}\theta(q'',t)\\
&= \frac{(-K\sin\alpha)}{2(2\pi)^{4d}} \iint d^dr d^dr' \int d^d\tilde{q} \iiint d^dq d^dq' d^dq''\ \tilde{\theta}(\tilde{q},t) J(q) \theta(q',t) \theta(q'',t) e^{ir(\tilde{q}+q+q' + q'')}e^{ir'(-q)}
\\
 &= \frac{(-K\sin\alpha)}{2(2\pi)^{3d}} \int d^dr  \int d^d\tilde{q} \iint  d^dq' d^dq''\ \tilde{\theta}(\tilde{q},t) J(0) \theta(q',t) \theta(q'',t) e^{ir(\tilde{q}+q' + q'')}  \\
  &= \frac{(-K\sin\alpha)}{2(2\pi)^{2d}} \iint d^d\tilde{q} d^dq\ \tilde{\theta}(\tilde{q},t) J(0) \theta(q,t) \theta(-\tilde{q}-q,t).
    \end{split}
\end{equation}
Here, note that all $J(0)$ parts are exactly canceled out among the three terms.

\medskip
In this regard, the remaining parts given by Eq.~(\ref{eq:third-term}) are
\begin{equation}
    \begin{split}
        \frac{\lambda_0}{(2\pi)^{2d+2}} \iint d^d\tilde{q} d^dq \iint d\omega d^d\tilde{\omega}\  \tilde{\theta}(\tilde{q},\tilde{\omega})|\tilde{q}|^\sigma\theta(q,\omega)\theta(-\tilde{q}-q,-\tilde{\omega}-\omega)\\
+\frac{g_0}{(2\pi)^{2d+2}} \iint d^d\tilde{q} d^dq \iint d\omega d^d\tilde{\omega}\   \tilde{\theta}(\tilde{q},\tilde{\omega})|q|^\sigma\theta(q,\omega)\theta(-\tilde{q}-q,-\tilde{\omega}-\omega),
    \end{split}
\end{equation}
where $\lambda_0=\frac{1}{2}c(\sigma,r_0)K\sin\alpha$ and $g_0=-c(\sigma,r_0)K\sin\alpha$.
In the above expressions, the Fourier transform with respect to $t$ has also been taken.

Finally, the last term $-(D/2)\tilde{\theta}^2$ gives $-\frac{D_0}{2} \tilde{\theta}(q,\omega)\tilde{\theta}(-q,-\omega)$.

\medskip
As a result, we obtain
\begin{equation}
    \begin{split}
        A[\tilde{\theta},\theta]&=\int_q\int_\omega \ \bigg[\tilde{\theta}(q,\omega)\{i\omega+\kappa_0|q|^\sigma\}\theta(-q,-\omega) \\
        &-\frac{D_0}{2} \tilde{\theta}(q,\omega)\tilde{\theta}(-q,-\omega) \\
        & +\lambda_0\int_{\tilde{q},{q}}\int_{\tilde{\omega},{\omega}}\tilde{\theta}(\tilde{q},\tilde{\omega})|\tilde{q}|^\sigma \theta(q,\omega) \theta(-\tilde{q}-q, -\tilde{\omega}-\omega)\\
        & +g_0\int_{\tilde{q},{q}}\int_{\tilde{\omega},{\omega}} \tilde{\theta}(\tilde{q},\tilde{\omega})|q|^\sigma \theta(q,\omega) \theta(-\tilde{q}-q, -\tilde{\omega}-\omega)\bigg],
    \end{split}
    \label{eq:S-MSR-momentum}
\end{equation}
where $\int_{q_1,q_2,...,q_n}=1/(2\pi)^{nd} \int d^dq_1 \int d^dq_2 ... \int d^dq_n$ and $\int_{\omega_1,\omega_2,...,\omega_n}=1/(2\pi)^{n} \int d\omega_1 \int d\omega_2 ... \int d\omega_n$.

\medskip
 
\section{C. Power counting}
To obtain the ``naive'' scaling behavior of each coupling with respect to the momentum scale $p$, we employ power counting. First, we define $\tilde{\chi}$ and $\chi'$ as the scaling exponents of the fields through $\tilde{\theta}'=b^{\tilde{\chi}}\tilde{\theta}$ and $\theta'=b^{\chi'} \theta$, under the scale transformations $q'=bq$ and $\omega'=b^z\omega$. One then readily obtains the following relation:
\begin{subequations}
    \begin{equation}
        D_0'=b^{d+2\chi'+3z}D_0,
    \end{equation}
    \begin{equation}
        \kappa_0'=b^{z-\sigma}\kappa_0,
    \end{equation}
    \begin{equation}
        \lambda_0'=b^{-(d+\chi'+\sigma)}\lambda_0,
    \end{equation}
    \begin{equation}
        g_0'=b^{-(d+\chi'+\sigma)}g_0,
    \end{equation}
    \label{eq:power-counting}
\end{subequations}
as the embedding dimension $d$, and $\tilde{\chi}+{\chi'}=-d-2z$ is imposed from the invariance of $\tilde{\theta}\partial_t \theta$ term.

\section{D. Derivation of the Wilson renormalization group equations}

In the following, we derive the renormalization-group (RG) equations for the model with the long-range kernel [Eqs.~(\ref{eq:RG-equations-Wilson}) in the main text] to one-loop order. We first illustrate the basic elements required to implement the Wilson RG~\cite{Wilson-Fisher-epsilon} in Fig.~\ref{fig:S1}. We emphasize the structural difference between our action and that of Ref.~\cite{Two-loop-KPZ}: there are two distinct three-point vertices, associated with $\lambda_0$ and $g_0$. As noted in the main text, these couplings must be treated separately, since the scaling exponents of $\tilde{\theta}$ and $\theta$ with respect to the momentum scale are different. Here, the subscript $0$ denotes bare coupling constants in the action. We will omit this subscript for the running couplings generated by the momentum-shell procedure in the Wilson RG.

\begin{figure*}[t]
    \centering
        \includegraphics[width=0.6\linewidth]{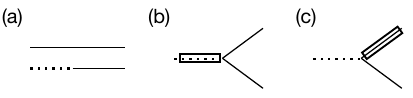}
\captionsetup{justification=raggedright,singlelinecheck=false}
    \caption{Basic diagrams of the dynamical renormalization-group (DRG) method for noisy identical Kuramoto--Sakaguchi model [Eq.~(\ref{eq:model}) in the main text]. (a) Correlation propagator $C_0(q,\omega)$ and response propagator $G_0(q,\omega)$. Three-point vertex diagrams for (b) $\lambda_0$ and (c) $g_0$. Here, each dotted line represents the $\tilde{\theta}$ field, whereas each solid line represents the $\theta$ field. The open square is attached to indicate the field carrying the momentum associated with $|q|^\sigma$. Comparison with the corresponding diagrams in Ref.~\cite{Two-loop-KPZ} highlights the differences.}
    \label{fig:S1}
\end{figure*}

\medskip
The terms that renormalize the couplings in $A_0$ (or the ``two-point vertex function''~\cite{Two-loop-KPZ,Tauber-PRX}) are represented by the Feynman diagrams in Fig.~\ref{fig:S2}. We note that all diagrams shown in this paper satisfy causality~\cite{tauber2014critical,Two-loop-KPZ,Tauber-PRX,Martin1973MSR,Janssen1976LagrangeanClassicalFieldDynamics}. One can first see that both Figs.~\ref{fig:S2}(a) and \ref{fig:S2}(b) contribute to the renormalization of $D_0$. We draw Figs.~\ref{fig:S2}(a) and \ref{fig:S2}(b) separately because they lead to different momentum integrations. Specifically, Fig.~\ref{fig:S2}(a) corresponds to
\begin{equation}
    \begin{split}
    \frac{g_0^2 D_0^2}{(2\pi)^{2d+1}} &\iint d\Omega d^dk \ \bigg[\frac{|\frac{q}{2}+k|^\sigma}{-i(\frac{\omega}{2}+\Omega)+\kappa_0|\frac{q}{2}+k|^\sigma}\frac{|\frac{q}{2}+k|^\sigma}{i(\frac{\omega}{2}+\Omega)+\kappa_0|\frac{q}{2}+k|^\sigma} \\
    &\qquad \qquad \quad \cdot \frac{1}{-i(\frac{\omega}{2}-\Omega)+\kappa_0|\frac{q}{2}-k|^\sigma}\frac{1}{i(\frac{\omega}{2}-\Omega)+\kappa_0|\frac{q}{2}-k|^\sigma}\bigg],
    \end{split}
    \label{eq:D-first-moemntum-integral}
\end{equation}
and Fig.~\ref{fig:S2} (b) corresponds to 

\begin{equation}
    \begin{split}
    \frac{g_0^2 D_0^2}{(2\pi)^{2d+1}} &\iint d\Omega d^dk \ \bigg[\frac{|\frac{q}{2}+k|^\sigma}{-i(\frac{\omega}{2}+\Omega)+\kappa_0|\frac{q}{2}+k|^\sigma}\frac{1}{i(\frac{\omega}{2}+\Omega)+\kappa_0|\frac{q}{2}+k|^\sigma} \\
    & \qquad \qquad \quad \cdot \frac{1}{-i(\frac{\omega}{2}-\Omega)+\kappa_0|\frac{q}{2}-k|^\sigma}\frac{|\frac{q}{2}-k|^\sigma}{i(\frac{\omega}{2}-\Omega)+\kappa_0|\frac{q}{2}-k|^\sigma}\bigg].
    \end{split}
    \label{eq:D-second-moemntum-integral}
\end{equation}

However, these expressions reduce to exactly the same integral in the IR limit, $\omega=0$ and $q=0$. One can readily see that the same property holds for all momentum and frequency integrals considered in this paper. Therefore, to obtain the correct coefficient for each renormalization integral, it suffices to account only for the symmetry factor of each diagram, irrespective of the position of the $|q|^\sigma$ term (i.e., the open square in the figures) among the diagrams that survive in the IR limit.

Accordingly, the effective noise strength $D$ is obtained as follows:
\begin{equation}
    D=D_0\left[1 + K_d\Lambda^{d-\sigma}\frac{g_0^2 D_0}{\kappa_0^3}
    \delta l\right],
    \label{eq:S-D-CG}
\end{equation}
when setting $\omega=0$ and $q=0$, applying the residue theorem to the internal frequency $\Omega$, and coarse-graining the modes within the momentum shell $k\in[\Lambda/b,\Lambda]$ with $b=e^{\delta l}>1$, where $\Lambda$ is the ultraviolet (UV) cutoff and $K_d\equiv S_d/[2(2\pi)^d]$.

An effective constant $\kappa$ of the propagator is renormalized by Figs.~\ref{fig:S2} (c), (d), and (e). First, Fig.~\ref{fig:S2} (c) represents
\begin{equation}
\begin{split}
    \frac{4D_0}{(2\pi)^{2d+1}}&\iint d\Omega d^dk \ \bigg[\frac{1}{-i(\frac{\omega}{2}-\Omega)+\kappa_0|\frac{q}{2}-k|^\sigma}\frac{1}{i(\frac{\omega}{2}-\Omega)+\kappa_0|\frac{q}{2}-k|^\sigma} \\
    &\qquad \qquad \quad \cdot \frac{|\frac{q}{2}+k|^\sigma}{-i(\frac{\omega}{2}+\Omega)+\kappa_0|\frac{q}{2}+k|^\sigma}\bigg],
\end{split}
\end{equation}
multiplied $g_0^2$. As in the renormalization of $D_0$, the momentum and frequency integrals for Figs.~\ref{fig:S2}(d) and \ref{fig:S2}(e), with couplings $\lambda_0 g_0$ and $\lambda_0^2$, respectively, become identical in the IR limit. Therefore, $\kappa_0$ is renormalized as
\begin{equation}
    \kappa=\kappa_0\left[1 - 2K_d\Lambda^{d-\sigma}\frac{(g_0^2+g_0\lambda_0+\lambda_0^2) D_0}{\kappa_0^3}\delta l\right].
    \label{eq:S-nu-CG}
\end{equation}

\begin{figure}[t]
    \centering
        \includegraphics[width=0.75\linewidth]{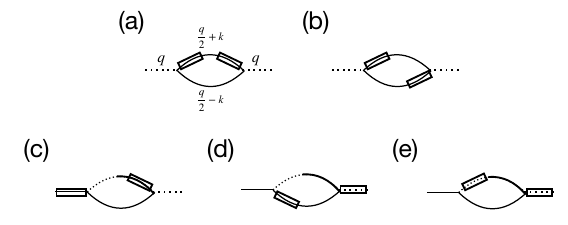}\captionsetup{justification=raggedright,singlelinecheck=false}
    \caption{Feynman diagrams representing the renormalization of the two-point vertex functions. Panels (a) and (b) contribute to the renormalization of $D_0$, whereas panels (c), (d), and (e) contribute to that of $\kappa_0$. Note that the momentum dependence of the field lines in (a) is also identical for the other diagrams.}
    \label{fig:S2}
\end{figure}

We depict the Feynman diagrams contributing to the renormalization of the couplings in $A_{\mathrm{int}}$ (or the ``three-point vertex function'') in Figs.~\ref{fig:S3} and \ref{fig:S4}. Specifically, the diagrams for the renormalization of $\lambda_0$ are shown in Fig.~\ref{fig:S3}, and in the IR limit the integral corresponding to each diagram is identical to that of
\begin{equation}
\begin{split}
    \frac{D_0}{(2\pi)^{2d+1}}&\iint d\Omega d^dk\ \bigg[ \frac{|k|^\sigma}{-i\Omega+\kappa_0|k|^\sigma}\frac{|\frac{q}{2}-k|^\sigma}{-i(\frac{\omega}{2}-\Omega)+\kappa_0|\frac{q}{2}-k|^\sigma}\\
    &\qquad \qquad \ \ \cdot \frac{1}{-i(\frac{\omega}{2}+\Omega)+\kappa_0|\frac{q}{2}+k|^\sigma}\frac{1}{i(\frac{\omega}{2}+\Omega)+\kappa_0|\frac{q}{2}+k|^\sigma} \bigg],
\end{split}
\end{equation}
multiplied by the corresponding coupling constants.

Specifically, $\lambda_0$ is renormalized by
\begin{equation}
    \lambda=\lambda_0\left[1 + 6K_d\Lambda^{d-\sigma}\frac{(g_0^2+g_0\lambda_0+\lambda_0^2) D_0}{\kappa_0^3}\delta l\right].
    \label{eq:S-lambda-CG}
\end{equation}

Applying a similar procedure to Fig.~\ref{fig:S4}, we find that $g_0$ is renormalized as
\begin{equation}
    g=g_0\left[1 +6K_d\Lambda^{d-\sigma}\frac{(g_0^2+g_0\lambda_0) D_0}{\kappa_0^3}\delta l\right].
    \label{eq:S-g-CG}
\end{equation}
Note that the term renormalizing $g_0$ does not contain a contribution of $g_0\lambda_0^2$, since such a term is geometrically forbidden.

\begin{figure}[t]
    \centering        \includegraphics[width=0.65\linewidth]{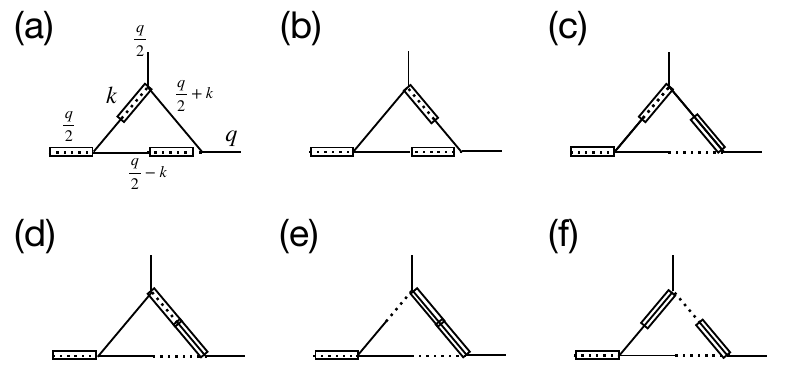}\captionsetup{justification=raggedright,singlelinecheck=false}
    \caption{Feynman diagrams representing the renormalization of $\lambda_0$. Note that the momentum dependence of the field lines in (a) is also identical for the other diagrams.}
    \label{fig:S3}
\end{figure}

\medskip
In addition to the coarse-graining steps in Eqs.~(\ref{eq:S-D-CG}), (\ref{eq:S-nu-CG}), (\ref{eq:S-lambda-CG}), and (\ref{eq:S-g-CG}), we perform the rescaling transformation with $b=e^{\delta l}>1$ to complete the Wilson RG procedure. Applying the rescaling relations in Eqs.~(\ref{eq:power-counting}), we obtain the RG equations for $\delta l\ll 1$ as follows:
\begin{subequations}
\begin{equation}
    \frac{dD}{dl}=\left[(d+2\chi'+3z)+K_d\Lambda^{d-\sigma}\frac{g^2 D}{\kappa^3}\right]D,
\end{equation}
\begin{equation}
    \frac{d\kappa}{dl}= \left[(z-\sigma)-2\frac{K_d}{\kappa^3}\Lambda^{d-\sigma}{D(g^2+\lambda^2+g\lambda)}\right]\kappa,
\end{equation}
\begin{equation}
    \frac{d\lambda}{dl}=\left[-(d+\chi'+\sigma) + 6K_d\Lambda^{d-\sigma}\frac{D}{\kappa^3}\left\{g^2 +\lambda g +\lambda^2\right\}\right]\lambda,
\end{equation}
\begin{equation}
    \frac{dg}{dl}=\left[-(d+\chi'+\sigma)+6K_d\Lambda^{d-\sigma}\frac{D}{\kappa^3} \left\{ g^2 +\lambda g \right\} \right]g,
\end{equation}
\end{subequations}
which are the same as Eqs.~(\ref{eq:RG-equations-Wilson}) in the main text with $\Lambda=1$.

To analyze the critical behavior, it is convenient to introduce the following dimensionless coupling constants:
\begin{equation}
    \tilde{g}_1 = \frac{g^2D}{\kappa^3}, \ \tilde{g}_2 = \frac{\lambda g D}{\kappa^3},\ \tilde{g}_3 = \frac{\lambda^2  D}{\kappa^3}.
    \label{eq:dimensionless-g}
\end{equation}
With Eq.~(\ref{eq:dimensionless-g}), the RG equations become
\begin{subequations}
    \begin{equation}
        \frac{d\tilde{g}_1}{dl}
=\tilde{g}_1\left[(-d+\sigma)+K_d\Lambda^{d-\sigma}\left\{19\tilde{g}_1 + 18\tilde{g}_2 +6 \tilde{g}_3\right\} \right],
    \end{equation}
    \begin{equation}
       \frac{d\tilde{g}_2}{dl}
=\tilde{g}_2\left[(-d+\sigma)+K_d\Lambda^{d-\sigma}\left\{19\tilde{g}_1 + 18\tilde{g}_2+12\tilde{g}_3\right\} \right],
    \end{equation}
    \begin{equation}
        \frac{d\tilde{g}_3}{dl}
=\tilde{g}_3\left[(-d+\sigma)+K_d\Lambda^{d-\sigma}\left\{19\tilde{g}_1 +18\tilde{g}_2+18\tilde{g}_3\right\} \right].
    \end{equation}
    \label{eq:S-RG-eqs}
\end{subequations}

\begin{figure}[b]
    \centering
        \includegraphics[width=0.45\linewidth]{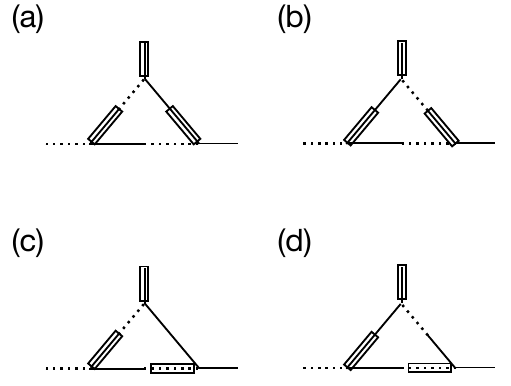}\captionsetup{justification=raggedright,singlelinecheck=false}
    \caption{Feynman diagrams representing the renormalization of $g_0$. Note that the momentum dependence of the field lines in (a) is also identical for the other diagrams.}
    \label{fig:S4}
\end{figure}

\section{E. Nontrivial fixed points}
Eqs.~(\ref{eq:S-RG-eqs}) have multiple fixed points. First, the trivial fixed point is
\begin{equation}
    \left(\tilde{g}_1, \tilde{g}_2, \tilde{g}_3\right)=\left(0,0,0\right),
    \label{eq:fp-trivial}
\end{equation}
and its linear stability is described by the eigenvalue $(-d+\sigma)$ for all directions of $\tilde{g}_1, \tilde{g}_2$, and $\tilde{g}_3$. That is, it is a stable fixed point in $\sigma<d$.

The nontrivial fixed points are given by
\begin{equation}
    \left(\tilde{g}_1, \tilde{g}_2, \tilde{g}_3\right)=\left(0,0,\frac{\Lambda^{\sigma-d}}{18K_d}(d-\sigma)\right),
    \label{eq:fp-nontrivial1}
\end{equation}
and
\begin{equation}
    \left(\tilde{g}_1, \tilde{g}_2, \tilde{g}_3\right)=\left(\tilde{g}_1^*,\frac{\Lambda^{\sigma-d}}{18K_d}\left[(d-\sigma)-19K_d\Lambda^{d-\sigma}\tilde{g}_1^*\right],0\right).
    \label{eq:fp-nontrivial2-whole}
\end{equation}
The \textit{physically admissible} nontrivial fixed point obtained from Eq.~(\ref{eq:fp-nontrivial2-whole}) is given by
\begin{equation}
    \left(\tilde{g}_1, \tilde{g}_2, \tilde{g}_3\right)=\left(\frac{\Lambda^{\sigma-d}}{19K_d}(d-\sigma),0,0\right),
    \label{eq:fp-nontrivial2}
\end{equation}
as described in the main text.

For $\sigma<d$, the first nontrivial fixed point in Eq.~(\ref{eq:fp-nontrivial1}) is unstable, with eigenvalue $(d-\sigma)$ along the eigenvector $(0,0,1)$. In contrast, it is stable along the eigenvectors $(-30,0,19)$ and $(0,-4,3)$, with eigenvalues $2(-d+\sigma)/3$ and $(-d+\sigma)/3$, respectively. Hence, these two eigenvectors locally define the ``critical manifold'' for the RG trajectories within the linearized analysis.

The second nontrivial fixed point in Eq.~(\ref{eq:fp-nontrivial2}) is unstable, with eigenvalue $(d-\sigma)$ along the eigenvector $(1,0,0)$; however, it is marginal at linear order along the other two eigenvectors $(-6,0,19)$ and $(-18,19,0)$. Therefore, its stability cannot be determined within the linear analysis. In this regard, we employ the center manifold theory in the following.

\section{F. Center manifold theory to determine stability of Eq.~(\ref{eq:fp-nontrivial2})}
For the linearly marginal fixed point in Eq.~(\ref{eq:fp-nontrivial2}), we first expand $\tilde{g}_1$, $\tilde{g}_2$, and $\tilde{g}_3$ around the fixed point as
\begin{equation}
    \begin{split}
        &\tilde{g}_1=\tilde{g}_1^*+\delta\tilde{g}_1, \\
        &\tilde{g}_2=\delta\tilde{g}_2,\\
         &\tilde{g}_3=\delta\tilde{g}_3,
    \end{split}
    \label{eq:expansion-ntr2}
\end{equation}
where $\tilde{g}_1^*={\Lambda^{\sigma-d}}(d-\sigma)/(19K_d)$. In the following, we set $\Lambda=1$, since it does not affect the stability, for convenience. The RG equations then become
\begin{subequations}
    \begin{equation}
        \begin{split}
        \frac{d(\delta \tilde{g}_1)}{dl}&= (\tilde{g}_1^*+\delta \tilde{g}_1)\left[(-d+\sigma)+K_d\left\{19(\tilde{g}_1^* +\delta \tilde{g}_1)+18\delta \tilde{g}_2 +6\delta \tilde{g}_3\right\}\right]\\
        &=\left(19K_d(\tilde{g}_1^*)^2+(-d+\sigma)\tilde{g}_1^* \right) + \delta \tilde{g}_1\left(38K_d\tilde{g}_1^* + (-d+\sigma) \right)\\
        &\qquad + K_d(18\delta \tilde{g}_2 + 6\delta \tilde{g}_3) \tilde{g}_1^* +K_d\delta \tilde{g}_1\left(19\delta \tilde{g}_1+18\delta \tilde{g}_2 + 6\delta \tilde{g}_3\right)\\
        &= \frac{(d-\sigma)}{19}\left[19\delta \tilde{g}_1 +18\delta \tilde{g}_2 +6\delta \tilde{g}_3 \right]+\delta \tilde{g}_1\left[K_d\left[19\delta \tilde{g}_1 + 18\delta \tilde{g}_2 +6\delta \tilde{g}_3\right]\right],
    \end{split}
\end{equation}
    \begin{equation}
        \begin{split}
            \frac{d(\delta \tilde{g}_2)}{dl} &= (\delta \tilde{g}_2)\left[(-d+\sigma)+K_d\left\{19(\tilde{g}_1^* +\delta \tilde{g}_1)+18\delta \tilde{g}_2 + 12\delta \tilde{g}_3\right\}\right]\\
            &= K_d\delta \tilde{g}_2 \left(19\delta \tilde{g}_1 + 18\delta \tilde{g}_2 + 12\delta \tilde{g}_3\right),
        \end{split}
    \end{equation}
    \begin{equation}
        \begin{split}
            \frac{d(\delta \tilde{g}_3)}{dl} &= (\delta \tilde{g}_3)\left[(-d+\sigma)+K_d\left\{19(\tilde{g}_1^* +\delta \tilde{g}_1)+18\delta \tilde{g}_2 + 18\delta \tilde{g}_3\right\}\right]\\
            &= K_d\delta \tilde{g}_3 \left(19\delta \tilde{g}_1 + 18\delta \tilde{g}_2 + 18\delta \tilde{g}_3\right).
        \end{split}
    \end{equation}
\end{subequations}
Secondly, we perform the following change of variables from $(\delta \tilde{g}_1,\delta \tilde{g}_2,\delta \tilde{g}_3)^\mathrm{T}$ to $(\delta\tilde{g}_1', \delta \tilde{g}_2',\delta\tilde{g}_3')^\mathrm{T}$:
\begin{equation}
    \begin{pmatrix} \delta \tilde{g}_1\\ \delta \tilde{g}_2 \\\delta \tilde{g}_3 \end{pmatrix} = \begin{pmatrix} 1 & -18& -6 \\ 0 & 19 & 0 \\ 0 & 0 & 19\end{pmatrix}
\begin{pmatrix} \delta \tilde{g}_1'\\ \delta \tilde{g}_2' \\\delta \tilde{g}_3' \end{pmatrix}=\begin{pmatrix} \delta \tilde{g}_1' -18 \delta \tilde{g}_2'-6\delta \tilde{g}_3' \\ 19\delta \tilde{g}_2'\\
19\delta \tilde{g}_3'\end{pmatrix}.
\end{equation}
RG equations then become
\begin{subequations}
    \begin{equation}
    \begin{split}
    \frac{d(\delta \tilde{g}_1)}{dl}&=\frac{d(\delta \tilde{g}_1')}{dl}-18\frac{d(\delta \tilde{g}_2')}{dl} -6\frac{d(\delta \tilde{g}_3')}{dl}\\
&=(d-\sigma)\delta \tilde{g}_1' +19K_d \delta \tilde{g}_1'\left[\delta\tilde{g}_1'-18\delta\tilde{g}_2'-6\delta\tilde{g}_3'\right],
    \label{eq:delta-g1-p}
    \end{split}
\end{equation}
    \begin{equation}
        \frac{d(\delta \tilde{g}_2')}{dl}=19K_d\left(\delta \tilde{g}_2' \delta \tilde{g}_1'+6\delta \tilde{g}_2'\delta \tilde{g}_3'\right),
        \label{eq:delta-g2-p}
    \end{equation}
    \begin{equation}
        \frac{d(\delta \tilde{g}_3')}{dl}=19K_d\left(\delta \tilde{g}_3' \delta \tilde{g}_1'+12{\delta \tilde{g}_3'}^2\right).
        \label{eq:delta-g3-p}
    \end{equation}
\end{subequations}
Here, the center manifold associated with the two ``slow'' variables $\tilde{g}_2'$ and $\tilde{g}_3'$ of this system is given by
\begin{equation}
    \delta \tilde{g}_1' = F(\delta \tilde{g}_2', \delta \tilde{g}_3'),
\end{equation}
which is zero when $\delta \tilde{g}_2'=\delta \tilde{g}_3'=0$.

We then write the formula of $F$ by
\begin{equation}
     F(\delta \tilde{g}_2', \delta \tilde{g}_3')= c_1{\delta \tilde{g}_2'}^2 + c_2{\delta \tilde{g}_2'}{\delta \tilde{g}_3'} + c_3 {\delta \tilde{g}_3'}^2 + O(\delta g_i' \delta g_j' \delta g_k'), 
     \label{eq:center-manifold-F}
\end{equation}
for $\left(i,j,k\in \{2,3\}\right)$. From Eq.~(\ref{eq:center-manifold-F}), we obtain
\begin{equation}
    \frac{d(\delta \tilde{g}_1')}{dl}=2c_1\delta \tilde{g}_2' \frac{d(\delta \tilde{g}_2')}{dl} + c_2\delta \tilde{g}_3'\frac{d(\delta \tilde{g}_2')}{dl} 
+ c_2 \delta \tilde{g}_2' \frac{d(\delta \tilde{g}_3')}{dl} + 2c_3 \delta \tilde{g}_3' \frac{d(\delta \tilde{g}_3')}{dl} .
\label{eq:d-deltag1-dl}
\end{equation}

Substituting Eqs.~(\ref{eq:delta-g2-p})--(\ref{eq:d-deltag1-dl}) into Eq.~(\ref{eq:delta-g1-p}), we obtain ${d(\delta \tilde{g}_1')}/{dl}$, represented by
\begin{equation}
    \begin{split}
        &2c_1\delta \tilde{g}_2' 19K_d\left(\delta \tilde{g}_2' \left( c_1{\delta \tilde{g}_2'}^2 + c_2{\delta \tilde{g}_2'}{\delta \tilde{g}_3'} + c_3 {\delta \tilde{g}_3'}^2 \right)+6\delta \tilde{g}_2'\delta \tilde{g}_3'\right) \\
&+ c_2 \delta \tilde{g}_3'19K_d\left(\delta \tilde{g}_2'\left( c_1{\delta \tilde{g}_2'}^2 + c_2{\delta \tilde{g}_2'}{\delta \tilde{g}_3'} + c_3 {\delta \tilde{g}_3'}^2 \right)+6\delta \tilde{g}_2'\delta \tilde{g}_3'\right)\\
&+c_2 \delta \tilde{g}_2'19K_d\left(\delta \tilde{g}_3' \left( c_1{\delta \tilde{g}_2'}^2 + c_2{\delta \tilde{g}_2'}{\delta \tilde{g}_3'} + c_3 {\delta \tilde{g}_3'}^2 \right)+12(\delta \tilde{g}_3')^2\right)\\
&+2c_3\delta \tilde{g}_3'19K_d\left(\delta \tilde{g}_3' \left( c_1{\delta \tilde{g}_2'}^2 + c_2{\delta \tilde{g}_2'}{\delta \tilde{g}_3'} + c_3 {\delta \tilde{g}_3'}^2 \right)+ 12(\delta \tilde{g}_3')^2\right),
    \end{split}
    \label{eq:LHS-center-manifold}
\end{equation}
is equal to 
\begin{equation}
    \begin{split}
        &(d-\sigma)\left(c_1\delta \tilde{g}_2'^2 + c_2\delta \tilde{g}_2' \delta \tilde{g}_3' + c_3 \delta \tilde{g}_3'^2 \right)+19K_d \left(c_1\delta \tilde{g}_2'^2 + c_2\delta \tilde{g}_2' \delta \tilde{g}_3' + c_3 \delta \tilde{g}_3'^2 \right)^2 \\
&-342K_d\delta \tilde{g}_2' \left(c_1\delta \tilde{g}_2'^2 + c_2\delta \tilde{g}_2' \delta \tilde{g}_3' + c_3 \delta \tilde{g}_3'^2 \right)-114K_d\delta \tilde{g}_3'\left(c_1\delta \tilde{g}_2'^2 + c_2\delta \tilde{g}_2' \delta \tilde{g}_3' + c_3 \delta \tilde{g}_3'^2 \right)\\
&+ 342K_d \left(\delta \tilde{g}_2' \left( c_1\delta \tilde{g}_2'^2 + c_2\delta \tilde{g}_2' \delta \tilde{g}_3' + c_3 \delta \tilde{g}_3'^2\right)+6\delta \tilde{g}_2'\delta \tilde{g}_3'\right)\\
&+114K_d\left(\delta \tilde{g}_3'\left( c_1\delta \tilde{g}_2'^2 + c_2\delta \tilde{g}_2' \delta \tilde{g}_3' + c_3 \delta \tilde{g}_3'^2\right)+12(\delta \tilde{g}_3')^2\right),
    \end{split}
    \label{eq:RHS-center-manifold}
\end{equation}
with Eq.~(\ref{eq:delta-g1-p}).

Before solving the equation using Eqs.~(\ref{eq:LHS-center-manifold}) and (\ref{eq:RHS-center-manifold}), we note that $c_1=0$, since it cannot be canceled by any other term. Therefore, Eq.~(\ref{eq:LHS-center-manifold}) reduces to
\begin{equation}
    \begin{split}
        &c_2 \delta \tilde{g}_3'19K_d\left(\delta \tilde{g}_2'\left(  c_2{\delta \tilde{g}_2'}{\delta \tilde{g}_3'} + c_3 {\delta \tilde{g}_3'}^2 \right)+6\delta \tilde{g}_2'\delta \tilde{g}_3'\right)\\
&+c_2 \delta \tilde{g}_2'19K_d\left(\delta \tilde{g}_3' \left(  c_2{\delta \tilde{g}_2'}{\delta \tilde{g}_3'} + c_3 {\delta \tilde{g}_3'}^2 \right)+12(\delta \tilde{g}_3')^2\right)\\
&+2c_3\delta \tilde{g}_3'19K_d\left(\delta \tilde{g}_3' \left(c_2{\delta \tilde{g}_2'}{\delta \tilde{g}_3'} + c_3 {\delta \tilde{g}_3'}^2 \right)+ 12(\delta \tilde{g}_3')^2\right),
    \end{split}
    \label{eq:result2-LHS}
\end{equation}
and Eq.~(\ref{eq:RHS-center-manifold}) becomes
\begin{equation}
    \begin{split}
        &(d-\sigma)\left( c_2\delta \tilde{g}_2' \delta \tilde{g}_3' + c_3 \delta \tilde{g}_3'^2 \right)+19K_d \left( c_2\delta \tilde{g}_2' \delta \tilde{g}_3' + c_3 \delta \tilde{g}_3'^2 \right)^2 \\
&-342K_d\delta \tilde{g}_2' \left( c_2\delta \tilde{g}_2' \delta \tilde{g}_3' + c_3 \delta \tilde{g}_3'^2 \right)-114K_d\delta \tilde{g}_3'\left(  c_2\delta \tilde{g}_2' \delta \tilde{g}_3' + c_3 \delta \tilde{g}_3'^2 \right)\\
&+ 342K_d\left(\delta \tilde{g}_2' \left(  c_2\delta \tilde{g}_2' \delta \tilde{g}_3' + c_3 \delta \tilde{g}_3'^2\right)+6\delta \tilde{g}_2'\delta \tilde{g}_3'\right)\\
&+114K_d\left(\delta \tilde{g}_3'\left( c_2\delta \tilde{g}_2' \delta \tilde{g}_3' + c_3 \delta \tilde{g}_3'^2\right)+12(\delta \tilde{g}_3')^2\right),
    \end{split}
    \label{eq:result2-RHS}
\end{equation}

By solving the equation obtained by setting Eq.~(\ref{eq:result2-LHS}) equal to Eq.~(\ref{eq:result2-RHS}), we obtain
\begin{equation}
    c_2(d-\sigma)=-2052K_d, \quad
c_3(d-\sigma)=-1368K_d,
\end{equation}
which gives $c_2=-2052 K_d/(d-\sigma)$ and $c_3=-1368K_d/(d-\sigma)$ with $c_1=0$.

Hence, the center manifold of our system is
\begin{equation}
    \delta \tilde{g}_1'= -\frac{12K_d}{(d-\sigma)}\left[(171){\delta \tilde{g}_2'}{\delta \tilde{g}_3'}+(114) {\delta \tilde{g}_3'}^2 \right],
    \label{eq:result-center-manifold}
\end{equation}
where the higher order terms are ignored.

With Eq.~(\ref{eq:result-center-manifold}), Eqs.~(\ref{eq:delta-g2-p}) and (\ref{eq:delta-g3-p}) become
\begin{subequations}
    \begin{equation}
    \frac{d(\delta \tilde{g}_2')}{dl}=19K_d\left( -\frac{12K_d}{(d-\sigma)}\left[(171){\delta \tilde{g}_2'^2}{\delta \tilde{g}_3'} +(114) \delta \tilde{g}_2'{\delta \tilde{g}_3'}^2 \right]+6\delta \tilde{g}_2'\delta \tilde{g}_3'\right),
        \label{eq:result-d-delta-g2}
    \end{equation}
    \begin{equation}
    \frac{d(\delta \tilde{g}_3')}{dl}=19K_d\left( - \frac{12K_d}{(d-\sigma)}\left[(171){\delta \tilde{g}_2'}{\delta \tilde{g}_3'}^2 + (114) {\delta \tilde{g}_3'}^3 \right]+12{\delta \tilde{g}_3'}^2\right),
        \label{eq:result-d-delta-g3}
    \end{equation}
\end{subequations}
respectively.

To analyze the stability, we utilize the polar coordinates as $\delta \tilde{g}_2'= \rho\cos\phi,  \delta \tilde{g}_3'=\rho \sin\phi$:
\begin{equation}
    \begin{split}
      \frac{d(\rho^2)}{dl}&=2\rho \frac{d\rho}{dl} = \frac{d(\delta \tilde{g}_2'^2+\delta \tilde{g}_3'^2)}{dl}=2\delta \tilde{g}_2'\frac{d\delta \tilde{g}_2'}{dl}+2\delta \tilde{g}_3'\frac{d\delta \tilde{g}_3'}{dl}\\
&=38K_d\left( -\frac{12K_d}{(d-\sigma)}\left[(171){\delta \tilde{g}_2'^3}{\delta \tilde{g}_3'} +(114) \delta \tilde{g}_2'^2{\delta \tilde{g}_3'}^2 \right]+6\delta \tilde{g}_2'^2\delta \tilde{g}_3'\right)\\
&\quad +38K_d\left( - \frac{12K_d}{(d-\sigma)}\left[(171){\delta \tilde{g}_2'}{\delta \tilde{g}_3'}^3 + (114){\delta \tilde{g}_3'}^4 \right]+12{\delta \tilde{g}_3'}^3\right)\\
&=-38K_d\bigg[U(171\rho^4\cos^3\phi \sin\phi + 114\rho^4\cos^2\phi\sin^2\phi)-6\rho^3\cos^2\phi \sin\phi\\
&\quad \ + U(171\rho^4\cos\phi \sin^3\phi +114\rho^4\sin^4\phi)-12\rho^3\sin^3\phi\bigg]\\
&= -38K_d\left[ U\left(\frac{171\rho^4}{2}(\sin2\phi)+114\rho^4\sin^2\phi \right)-6\rho^3\sin\phi(1+\sin^2\phi) \right],
    \end{split}
\end{equation}
where $U\equiv 12K_d/(d-\sigma)>0$.

Thus, we obtain
\begin{equation}
    \frac{d\rho}{dl}=-19 K_d \rho^2\left[U\rho\left(\frac{171}{2}\sin2\phi+114\sin^2\phi\right)-6\sin\phi(1+\sin^2\phi)\right].
\end{equation}
Next, $d\phi/dl$ is obtained by the relation $\phi=\tan^{-1}(\delta \tilde{g}_3'/\delta \tilde{g}_2')$, which gives 
\begin{equation}
\begin{split}
    \frac{d\phi}{dl}&=\frac{1}{\rho^2}\left(\delta \tilde{g}_2'\frac{d(\delta \tilde{g}_3')}{dl} - \delta \tilde{g}_3'\frac{d(\delta \tilde{g}_2')}{dl} \right)\\
&=\frac{19K_d}{\rho^2}\bigg(\left(- U\left[171{\delta \tilde{g}_2'}^2{\delta \tilde{g}_3'}^2 + 114 \delta \tilde{g}_2'{\delta \tilde{g}_3'}^3 \right]+12\delta \tilde{g}_2'{\delta \tilde{g}_3'}^2\right)\\
&\qquad\qquad-\left( -U\left[171{\delta \tilde{g}_2'}^2{\delta \tilde{g}_3'}^2 +114 \delta \tilde{g}_2'{\delta \tilde{g}_3'}^3 \right]+6\delta \tilde{g}_2'{\delta \tilde{g}_3'}^2\right)\bigg)\\
&=\frac{1}{\rho^2}\left(114K_d\delta \tilde{g}_2' {\delta \tilde{g}_3'}^2\right),
\end{split}
\end{equation}
which results in  
\begin{equation}
    \frac{d\phi}{dl}=114K_d\left(\rho \cos\phi\sin^2\phi\right),
\end{equation}
for which $\phi^*=\pi/2$ is the only stable fixed point as $\rho>0$, while other ones are either unstable or a semi-side fixed point.

Substituting $\phi=\pi/2$ in the IR limit, we find
\begin{equation}
    \frac{d\rho}{dl}=-19K_d\rho^2\left[114U\rho - 12\right],
\end{equation}
which indicates that the nontrivial fixed point in Eq.~(\ref{eq:fp-nontrivial2}) is unstable along all three eigendirections, and therefore cannot affect the critical behavior. In other words, the critical manifold is determined solely by Eq.~(\ref{eq:fp-nontrivial1}), as depicted in Fig.~\ref{fig:2} of the main text.
\medskip

\section{G. Analytical curve of $\alpha_c$ as a function of $\sigma$}
As described in the previous sections, Eq.~(\ref{eq:fp-nontrivial1}) locally defines the critical manifold (see Fig.~\ref{fig:2} in the main text). Here, we explain how to determine the critical phase lag $\alpha_c$ as a function of $\sigma$. From Eq.~(\ref{eq:fp-nontrivial1}), we obtain the intersection point from the following critical surface:
\begin{equation}
    (0,0,\tilde{g}_3^*)+(-30,0,19)s_1 + (0,-4,3)s_2
    \label{eq:critical-surface}
\end{equation}
where $\tilde{g}_3^*=\Lambda^{\sigma-d}(d-\sigma)/(18K_d)$, and the following ``physical line'' of our bare parameters:
\begin{equation}
B(4,-2,1)
\end{equation}
where $B$ is a constant factor to represent the straight line through the origin (trivial fixed point in Eq.~(\ref{eq:fp-trivial})). 

Then, we can use
\begin{equation}
    -30 s_1=4B,\quad 4s_2=2B,\quad (19s_1+3s_2)+\tilde{g}_3^*=B,
\end{equation}
from which we obtain the value of $B$ at the intersection as $B=(30/61)\tilde{g}_3^* $. We then need only determine $\alpha_c$ such that
\begin{equation}
    \frac{\lambda^2_0 D_0}{\kappa_0^3}=\frac{D_0}{4Kc(\sigma,r_0)}\tan^2\alpha_c\sec\alpha_c=\frac{5\Lambda^{\sigma-d}}{183K_d}(d-\sigma).
    \label{eq:equation-intersection}
\end{equation}
Expanding the left-hand side of Eq.~(\ref{eq:equation-intersection}) for $\alpha_c\ll1$, we obtain the following closed-form expression for $\alpha_c$:
\begin{equation}
    |\alpha_c|\approx \sqrt{G\left[\frac{-\Gamma(-\sigma/2)}{\Gamma(\sigma/2)}(d-\sigma)\right]}
    \label{eq:alpha_c}
\end{equation}
where
\begin{equation}
    G\equiv  \frac{K}{D_0}\frac{20(r_0/2)^\sigma}{183K_d\Lambda^{d-\sigma}}.
    \label{eq:G-coeff}
\end{equation}

Note that the square of the right-hand side of Eq.~(\ref{eq:alpha_c}) is positive in the regime $0<\sigma <d$ with $d=1$ and 2, so that it yields a physical solution to elucidate phase transitions in Fig.~\ref{fig:1} of the main text. We also see that $|\alpha_c|$ increases as the coupling strength $K$ of the phase oscillators increases or as the noise strength $D_0$ decreases, which is consistent with physical intuition. Specifically, for a fixed $\alpha$, the system can sustain stable synchronization more easily when the interaction is stronger or the noise is weaker.

\medskip
\section{H. Numerical Phase diagram on complex networks}

As a complex network with tunable spectral dimension, we use the long-range random ring (LRRR)
network~\cite{LRRR}, whose spectral dimension can be continuously varied over $d_s\in[1,\infty)$. We briefly describe the LRRR network introduced in Ref.~\cite{LRRR}, as it is used in our additional numerical simulations [Fig.~\ref{fig:S5}] mentioned in the Discussion of the main text. The network consists of a ring backbone of $N$ regularly spaced nodes labeled by $i=1,\dots,N$ with periodic boundary conditions $\theta_{N+1}=\theta_1$. The distance between nodes $i$ and $j$ is defined as
\[
r_{ij}=\min(|i-j|,\,N-|i-j|),
\]
and a link $e_{ij}$ is present with probability
\[
p_{ij}=r_{ij}^{-(1+\sigma)}.
\]
For this network, the spectral dimension exceeds unity, $d_s>1$, when $\sigma<2$~\cite{LRRR}.

\begin{figure}[t]
    \centering
    \includegraphics[width=0.6\linewidth]{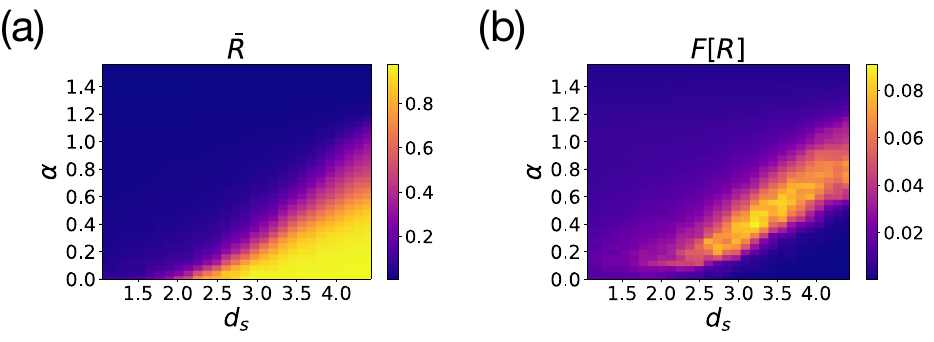}
    \captionsetup{justification=raggedright,singlelinecheck=false}
    \caption{Phase diagram of Eq.~(\ref{eq:model}) of the main text in the $(d_s,\alpha)$ plane, where $J_{ij}$ is replaced by the adjacency matrix defined on complex networks with a continuously tunable spectral dimension~\cite{LRRR}. The initial phases $\theta_i$ are independently drawn from a uniform distribution, with network size $N=2^{14}$ and noise strength $D=4\times10^{-2}$. The color bars show ensemble-averaged values over 20 network realizations of (a) the long-time average of the synchronization order parameter, $\bar{R}$ [Eq.~(\ref{eq:r}) of the main text], and (b) the dynamical fluctuation, $F[R]$ [Eq.~(\ref{eq:F[r]}) of the main text] obtained by numerical integration with $t_r=4.8\times10^4$ and $t_{\mathrm{avg}}=2\times10^3$.}
    \label{fig:S5}
\end{figure}

\medskip
To obtain the estimates of $d_s$ for \textit{finite} networks, we can see the following relation from Eq.~(\ref{eq:rho(lambda)}) in the main text:
\begin{equation}
    k\sim \rho_c(\lambda_k)\sim \lambda_k^{d_s/2},
    \label{eq:cumulant-dist-discrete}
\end{equation}
for the $k$th eigenvalue. Hence, we use $\lambda_k\sim k^{2/d_s}$.

We estimate the values of $d_s$ shown in Fig.~\ref{fig:S5} from the eigenvalue spectrum of $\mathbf{L}$ using Eqs.~(\ref{eq:rho(lambda)}) and (\ref{eq:normalized-L}). We compute the 50 smallest nonzero eigenvalues with the Spectra C++ library built on Eigen~\cite{spectra}, and apply a greedy algorithm that simultaneously determines an optimal fitting window and estimates $d_s$ by maximizing
\begin{equation}
    F_{\mathrm{score}}= 1-\frac{\sum_k\left(\log \lambda_k-f_\mathrm{pred}(k)\right)^2}{\sum_k \left(\log \lambda_k-\langle \log \lambda_k\rangle \right)^2},
    \label{eq:score-fn}
\end{equation}
where $\lambda_k$ is the $k$th eigenvalue of $\mathbf{L}$ and $f_{\mathrm{pred}}(k)=a\log k+b$. The spectral dimension is estimated as $d_s=2/a$. The algorithm is initialized at $k_0=1,2,3$, and $4$, and for each $k_0$ chooses the optimal window ending at $\lambda_{k'}$. The final estimate is the one that maximizes Eq.~(\ref{eq:score-fn}) over all such choices. Representative fits are shown in Figs.~\ref{fig:S6}.

\medskip
To construct the ensemble-averaged phase diagram in Fig.~\ref{fig:S5}, we also average the estimated values of $d_s$ over different realizations of the LRRR network and remap them to a uniformly spaced $d_s$ axis. This step is necessary because each realization of $\bar{R}$ and $F[R]$ is obtained on its own network geometry, so the $\sigma$ axis cannot be replaced naively by an averaged relation $d_s=f(\sigma)$. Even at fixed $\sigma$, different realizations generated from $p_{ij}$ generally yield different values of $d_s$, which may otherwise introduce artificial fluctuations near $\alpha_c$ when the data are plotted against $d_s$. 

We begin from a fixed list of $\sigma\in[0.3,1.25]$ with spacing $\Delta\sigma=0.025$. Because the corresponding values of $d_s$ are not uniformly distributed~\cite{LRRR}, we apply a greedy algorithm that minimizes the variance of successive $d_s$ gaps by removing selected pairs $\{d_s,\sigma\}$, thereby fixing the number of ticks on the $d_s$ axis. This yields a realization-dependent discretized set of $d_s$ values. We then perform linear interpolation over $d_s\in[d_s^m,d_s^M]$, where $d_s^m$ and $d_s^M$ are the minimum and maximum values obtained from the realizations, respectively, to produce the phase diagram as a function of $d_s$.

The numerical estimates of the spectral dimension $d_s$ used in Fig.~\ref{fig:S5} were obtained using the algorithm described above. Figs.~\ref{fig:S6} show examples of the estimation of $d_s$ from the eigenvalues of the normalized Laplacian for a single realization of the LRRR~\cite{LRRR} network. Each panel shows that even a single network realization yields a fitting function in good agreement with the eigenvalue data. We also observe that, at a given value of $\sigma$, the spectral dimension $d_s$ of the LRRR network differs from that of the fully connected graph with long-range weights $J_{ij}\propto(r_{ij}^2+r_0^2)^{-(1+\sigma)/2}$, as described in Ref.~\cite{LRRR}. For example, for a realization with $\sigma=0.875$ and $N=2^{14}$, we obtain the estimate $d_s\sim 2$ [Fig.~\ref{fig:S6}(b)], whereas the fully connected weighted graph in the thermodynamic limit gives $d_s=2/0.875$, which is clearly different from $2$ (see Eq.~(\ref{eq:d_s_annealed}) in the main text).

\begin{figure*}[!htbp]
    \centering
        \includegraphics[width=0.6\linewidth]{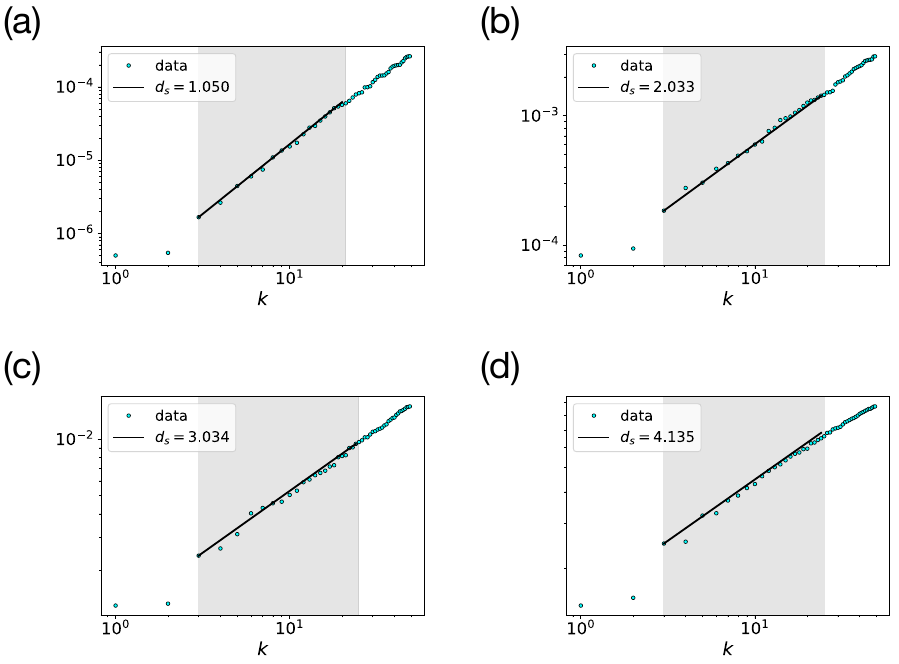}
\captionsetup{justification=raggedright,singlelinecheck=false}
    \caption{Examples illustrating the application of the algorithm used to obtain the numerical estimates of $d_s$. In each panel, the scatter points represent the $k$th eigenvalue $\lambda_k$ of the normalized Laplacian for a single realization of the LRRR network~\cite{LRRR} with $N=2^{14}$ nodes. The black solid line denotes the fit obtained only within the optimized window determined by the method described in Sec.~H; the fitted range is indicated by the gray region in each panel. The values of $\sigma$ in the edge connection probability $p_{ij}$ between nodes $i$ and $j$ are (a) $\sigma=1.225$, (b) $\sigma=0.875$, (c) $\sigma=0.65$, and (d) $\sigma=0.35$. The estimate of $d_s$ obtained for each value of $\sigma$ is shown in the legend of the corresponding panel. In Fig.~\ref{fig:S6}(d), all eigenvalues are of order $10^{-2}$. Note that, according to Eq.~(\ref{eq:cumulant-dist-discrete}), a larger slope corresponds to a smaller value of $d_s$.}
    \label{fig:S6}
\end{figure*}
\end{document}